\documentclass{IEEEtran}
\usepackage{iftex}
\ifPDFTeX
  \usepackage[T1]{fontenc}
\fi
\usepackage{amsmath,amssymb}
\usepackage{cite}
\usepackage{graphicx}
\usepackage{xcolor}
\usepackage[caption=0,font=footnotesize,subrefformat=parens]{subfig}
\usepackage{tikz}
\usetikzlibrary{arrows.meta,positioning}
\usepackage{array}
\usepackage{makecell}
\usepackage[colorlinks=true,linkcolor=black,citecolor=blue,urlcolor=blue]{hyperref}
\usepackage{microtype}
\usepackage{orcidlink}

\begin{document}

\title{VSWR-Resilient Mm-Wave and Cm-Wave PAs \\ for Large-Scale Phased Arrays}

\author{
Chenhao~Chu\orcidlink{0000-0002-2562-9599},~\IEEEmembership{Member,~IEEE,}
Masoud~Pashaeifar\orcidlink{0000-0002-5834-5461},~\IEEEmembership{Member,~IEEE,}\\
Filippo~Svelto,~\IEEEmembership{Graduate Student Member,~IEEE,}
and~Hua~Wang\orcidlink{0000-0003-4952-5505},~\IEEEmembership{Fellow,~IEEE}

\thanks{This work was sponsored in part by Mitsubishi Electric (Kanagawa, Japan), ETH Zurich internal grants, the HORIZON-JU-SNS-2023 ``6G-REFERENCE'' Project under Project 101139155, HORIZON-KDT-JU-2023-1-IA ``SOIL'' Project under Grant 101139785, Horizon-JU-SNS-2024 ``X-TREME 6G'' Project under Grant 101192681, and Swiss State Secretariat for Education, Research, and Innovation (SERI) through SwissChips Initiative. (\emph{Corresponding author: Chenhao Chu}).}%
\thanks{Chenhao Chu, Filippo Svelto, and Hua Wang are with the Department of Information Technology and Electrical Engineering (D-ITET), ETH Zürich, 8092 Z\"urich, Switzerland (e-mail: chenhao.chu@iis.ee.ethz.ch; wanghua@ethz.ch).}
\thanks{Masoud Pashaeifar is with NXP Semiconductors, the Netherlands (e-mail: masoud.pashaeifar@nxp.com).}
}%
\markboth{This manuscript is a preprint.}%
{This manuscript is a preprint.}
\maketitle
\bstctlcite{IEEEexample:BSTcontrol}

\begin{abstract}
Large-scale mm-Wave and cm-Wave phased arrays have become central to wireless communication and sensing systems, including terrestrial 5G/6G links and basestations, non-terrestrial network (NTN), satellite communication (SATCOM), radar, and relay applications. In dense arrays, the power amplifiers (PAs) and the antenna array interact with each other: the array pattern and radiation performance depend on the amplitude/phase of the PA output signals, while the load impedance experienced by each PA varies with frequency, scan angle, and element position. This active antenna impedance, commonly described as voltage standing wave ratio (VSWR) variation, is mainly caused by antenna inter-element mutual coupling and is further shaped by package/interconnect parasitics and the surrounding environment. As a result, each PA can significantly deviate from its optimum large-signal operating condition, degrading key large-signal metrics, including output power \(P_{\mathrm{out}}\), power gain (PG), power-added efficiency (PAE), and AM--AM/AM--PM, while also reducing reliability margin. These element-level variations further affect array EIRP consistency, EVM headroom, link budget, thermal density, and beamforming calibration, substantially complicating PA design for wideband, wide-scan-angle arrays. This review first introduces the origins of antenna VSWR in phased arrays and the antenna--PA interactions that connect the antenna reflection coefficient \(\Gamma_{\mathrm{ant}}\) and PA output matching \(S_{22}\) to delivered-power and transmitted-phase variation through the \(S_{22}\Gamma_{\mathrm{ant}}\) dependence. Reverse-coupled excitation and reverse intermodulation distortion (RIMD) are also introduced and discussed. This motivates PA designs that achieve simultaneous output and loadline matching (SOLM), enabling a low output reflection coefficient, i.e., small \(\lvert S_{22}\rvert\), without significantly compromising large-signal performance. Recent advances in mm-Wave and cm-Wave VSWR-resilient silicon integrated PA techniques and circuit demonstrations are then reviewed. Finally, future challenges and research opportunities toward compact, load-insensitive, energy-efficient, high-power-density, and calibration-scalable integrated PAs are discussed.
\end{abstract}


\section{Introduction}

\IEEEPARstart{L}{arge-scale} mm-Wave and cm-Wave phased arrays have become central to wireless communication and sensing systems, including terrestrial 5G/6G links, non-terrestrial satellite communication (SATCOM), radar, and relay applications. By enabling spatial power combining, electronic beam steering, and multi-element integration, phased arrays provide the effective isotropic radiated power (EIRP), coverage, link margin, and spatial selectivity required by modern terrestrial and non-terrestrial wireless systems~\cite{wang_millimeter-wave_2021,sadhu_28-ghz_2017,noauthor_phased_nodate,pashaeifar_chain-weaver_2024,hausmair_prediction_2017,dhar_reflection-aware_2018,fager_linearity_2019,capelli-mouvand_5g_2021,atanasov_reverse_2020,wang_surface-wave_2017,holzman_use_2013}. As the interface between the transceiver electronics and the antenna array, the power amplifier (PA) is a key front-end block~\cite{noauthor_pa_nodate,camarchia_review_2020,asbeck_power_2019,pashaeifar_millimeter-wave_2021}. It often dominates transmitter (TX) power consumption and nonlinear distortion while determining the available output power \(P_{\mathrm{out}}\), power-added efficiency (PAE), linearity, and reliability.

\begin{figure}[t!]
    \centering
    \includegraphics[width=0.95\linewidth]{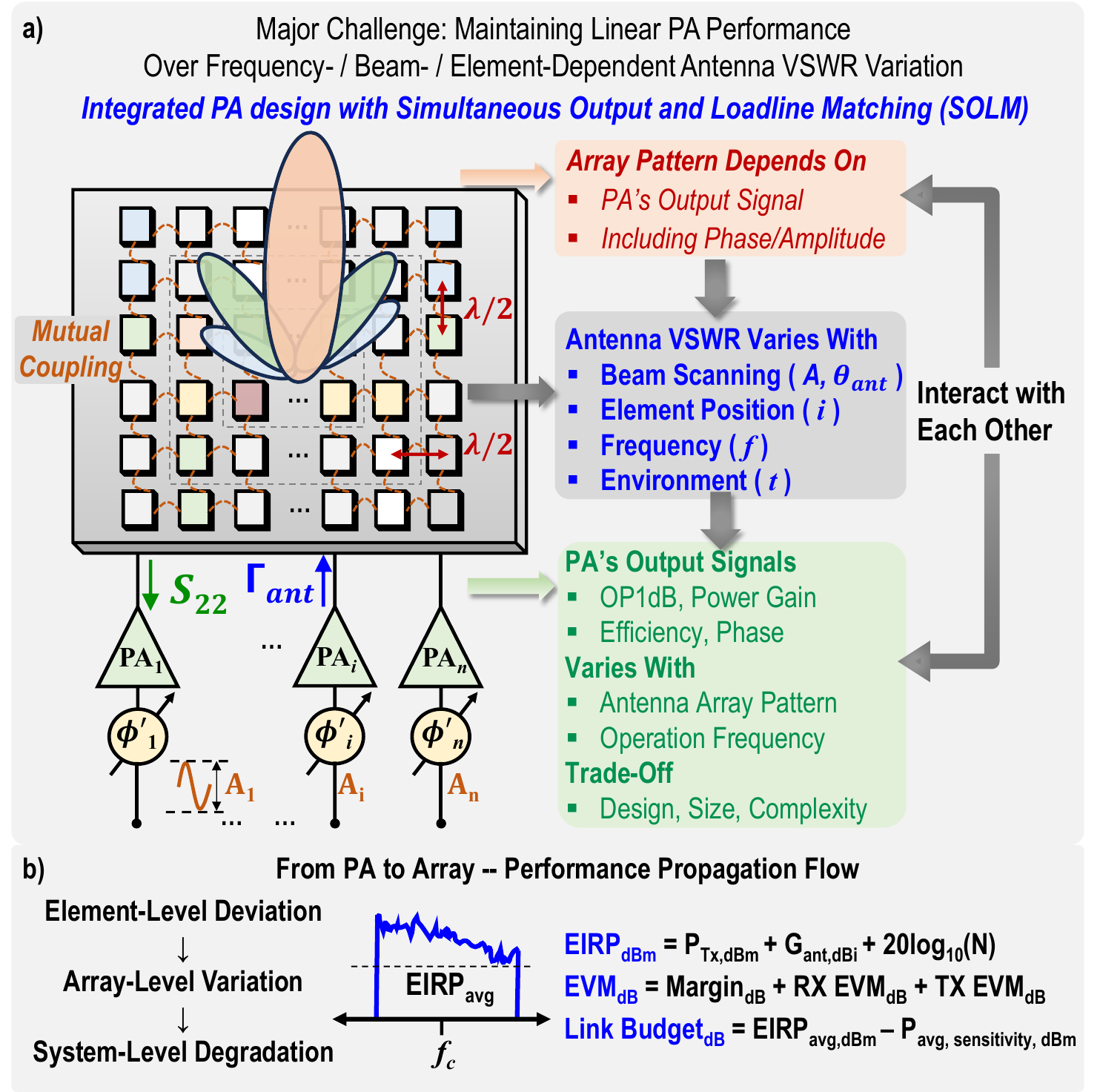}
    \caption{Antenna VSWR variation in large-scale phased arrays. (a) PA--array interaction through beam-, frequency-, element-position-, and environment-dependent load variation. (b) Performance propagation from element-level PA variations to array- and system-level metrics.}
    \label{fig:intro_vswr_overview}
    \vspace{-0.5em}
\end{figure}

In dense arrays, the PAs and the antenna array directly interact with each other, as shown in Fig.~\ref{fig:intro_vswr_overview}(a). The array pattern and radiation performance depend on the amplitude and phase of the PA output signals, while the load impedance experienced by each PA varies with frequency, scan angle, and element position. This antenna array active impedance variation is mainly caused by mutual coupling among closely spaced antenna elements and is further shaped by array excitation, package/interconnect parasitics, and the surrounding environment. It is also exacerbated by wide bandwidths and wide scan angles and is commonly described as antenna voltage standing wave ratio (VSWR) variation~\cite{munzer_broadband_2023,liu_2739-ghz_2025,kahn_active_1969,boryssenko_wave-based_2003,kalfa_fast_2013,chen_review_2018,pashaeifar_thesis_2024,liu_advanced_2025}.

Such frequency-, beam-, and element-position-dependent VSWR conditions change the PA operating condition and degrade PA large-signal performance. As shown in Fig.~\ref{fig:intro_vswr_overview}(b), the element-level PA variations then propagate to array- and system-level performance. Variations in \(P_{\mathrm{out}}\), \(\mathrm{OP}_{1\mathrm{dB}}\), and PG affect EIRP consistency and link budget; phase variation perturbs the intended beamforming weights and sidelobe levels; PAE degradation increases thermal density; and load-dependent AM--AM/AM--PM behavior reduces the available EVM and spectral-regrowth margin~\cite{pashaeifar_millimeter-wave_2021,pashaeifar_millimeter-wave_2026,pashaeifar_144_2021,pashaeifar_327_2024,eleraky_compact_2026,eleraky_55_2025,eleraky_204_2026,ghorbanpoor_332_2026}. Moreover, antenna VSWR poses a significant reliability risk to PAs, as large voltage swings can exceed transistor breakdown limits and cause permanent damage. Reverse-coupled signals from neighboring elements can also enter the PA output and mix with the locally generated signal, producing reverse intermodulation distortion (RIMD)~\cite{atanasov_reverse_2020}. This raises three fundamental PA design questions:
\begin{itemize}
\item[-] How to evaluate, quantify, and mitigate the PA's VSWR load sensitivity?
\item[-] Should the PA output matching \(S_{22}\) also be considered in the design process?
\item[-] Further, can high-efficiency PAs achieve simultaneous output matching \((S_{22})\) and loadline matching (\(Z_{\mathrm{opt}}\))?
\end{itemize}

Conventional radio-frequency (RF) PA design methodology treats the PA devices as RF current sources and primarily focuses on PA loadline matching (\(Z_{\mathrm{opt}}\)) to optimize PA \(P_{\mathrm{out}}\) and efficiency. Throughout this paper, \(S_{22}\) denotes the complex output reflection coefficient; when expressed in decibels, it denotes \(20\log_{10}\lvert S_{22}\rvert\). Output matching \(S_{22}\) is often used only for stability checking. \emph{Sometimes, poor output matching, i.e., \(S_{22}\approx-1\) to \(-5~\mathrm{dB}\), is even taken as a success for achieving high PA efficiency, since an ideal RF current source connected through a lossless output matching network should present \(\lvert S_{22}\rvert=1\), or \(S_{22}=0~\mathrm{dB}\).} Also, in practical standalone RF PAs, antenna load variations can be largely mitigated by magnetic isolators at the PA output. However, for high-density cm-Wave and mm-Wave phased arrays, employing output magnetic isolators for each PA becomes impractical. Thus, the PAs are directly subjected to antenna load variations described by \(\Gamma_{\mathrm{ant}}\), which interacts with the PA output reflection coefficient \(S_{22}\), introducing an \(S_{22}\Gamma_{\mathrm{ant}}\)-dependent sensitivity in delivered power and transmitted phase to antenna VSWR variation. This motivates array-aware PA designs to simultaneously achieve the output conjugate-match condition and loadline matching, without significantly compromising PA large-signal performance~\cite{pecile_study_2025,eleraky_compact_2026,eleraky_204_2026,eleraky_55_2025,capelli-mouvand_5g_2021}. Moreover, integrated PAs for large-scale arrays are required to maintain predictable large-signal performance over the expected antenna VSWR while satisfying compactness, power-density, and calibration-scalability constraints. Recent works have started to explore this topic through nonreciprocal
PA--antenna interfaces~\cite{pashaeifar_millimeter-wave_2026,ghorbanpoor_332_2026,dinc_millimeter-wave_2017,nagulu_fully-integrated_2018,nagulu_285_2019}, reflected-wave redistribution~\cite{pashaeifar_millimeter-wave_2021,pashaeifar_chain-weaver_2024,diverrez_24-31ghz_2023,ghorbanpoor_332_2026}, sensing-assisted
reconfiguration~\cite{munzer_broadband_2022,munzer_broadband_2023,munzer_single-ended_2021,munzer_single-ended_2022,liu_24_2025,bowers_integrated_2013}, and \emph{output-matched} PA design~\cite{eleraky_compact_2026,eleraky_204_2026,eleraky_55_2025}. Doherty and other load-modulated PAs are further considered where VSWR resilience must be combined with output power back-off (PBO) efficiency enhancement. Due to the limited scope, this paper will mainly focus on mm-Wave and cm-Wave silicon PAs, while compound semiconductor devices and some III--V PA examples will be discussed more briefly.

The paper is organized as follows. Section~II discusses antenna--PA interaction and details the impact of \(S_{22}\Gamma_{\mathrm{ant}}\) in VSWR-resilient PA design. Section~III reviews circuit and architecture techniques for recent VSWR-resilient PA examples. Section~IV summarizes representative mm-Wave and cm-Wave VSWR-resilient PA demonstrations. Section~V discusses future trends, challenges, and innovation opportunities, and Section~VI concludes the paper.

\section{Fundamentals of VSWR-Resilient PA Design}

In large-scale phased arrays, the PA and antenna element form a coupled interface. The PA output amplitude and phase set the array excitation, while the frequency-, scan-angle-, and element-position-dependent antenna impedance in turn perturbs the PA operating condition. This section discusses the system-level significance of VSWR resilience and the antenna--PA interactions that connect the antenna reflection coefficient \(\Gamma_{\mathrm{ant}}\), the PA output reflection coefficient \(S_{22}\), delivered power, transmitted phase, and RIMD. Small- and large-signal evaluation regimes are then discussed, followed by the relationship between PA output matching and optimum load-pull matching in VSWR-resilient PA design.

\subsection{System-Level Significance of VSWR Resilience}

PA designs are typically optimized for a nominal \(50~\Omega\) load. In practical wireless systems, however, the load presented to the PA often deviates from this value. In user equipment (UE), such as handsets, user proximity, platform motion, and nearby reflections can perturb the antenna load. In phased arrays, mutual coupling among the antenna elements produces active array impedance variations, and even well-designed arrays can exhibit VSWRs of approximately 3:1~\cite{pashaeifar_chain-weaver_2024, eleraky_compact_2026, wang_millimeter-wave_2021, munzer_broadband_2023, liu_2739-ghz_2025}.

\begin{figure}[t!]
    \centering
    \includegraphics[width=0.95\linewidth]{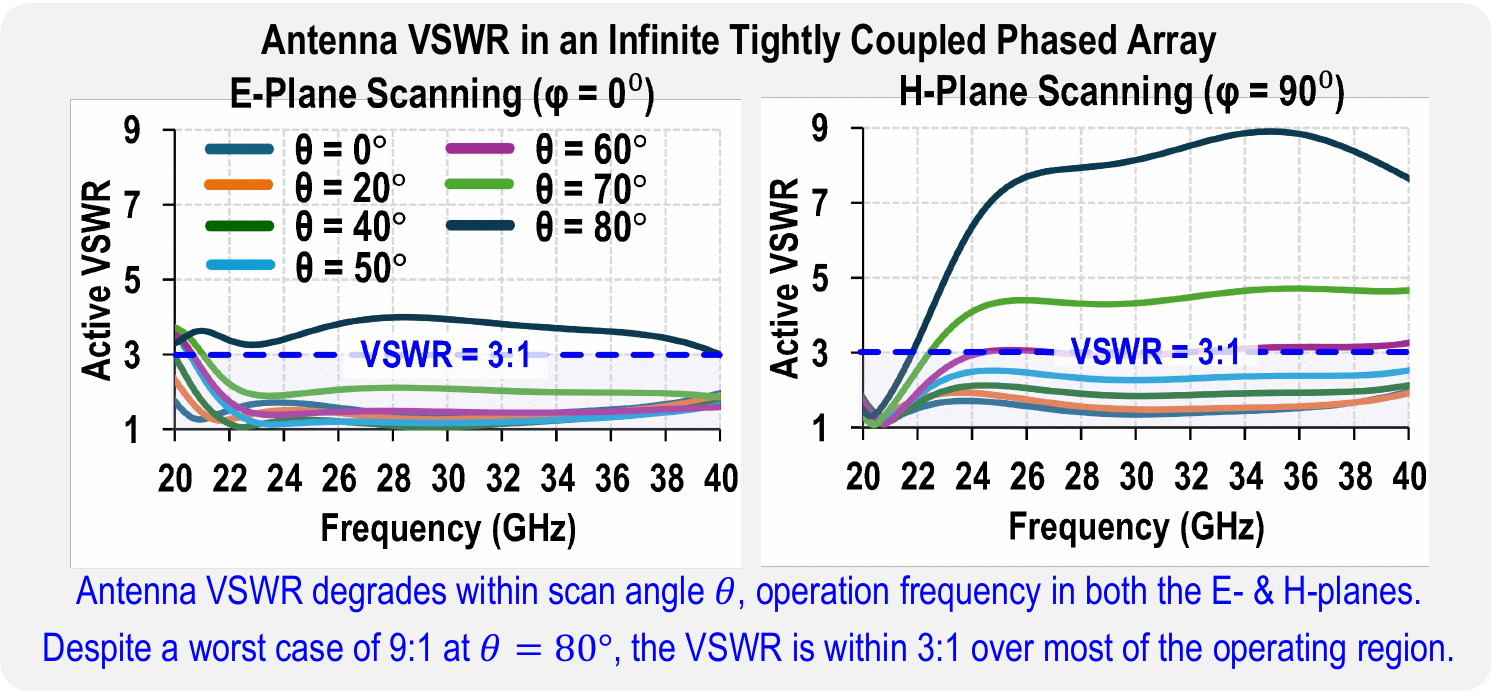}
    \caption{Simulated active VSWR of an infinite, tightly coupled phased array versus frequency and scan angle in the E- and H-planes~\cite{liu_2739-ghz_2025}.}
    \label{fig:antenna_vswr_infinite_array}
    \vspace{-1.0em}
\end{figure}

\begin{figure}[t!]
    \centering
    \includegraphics[width=0.95\linewidth]{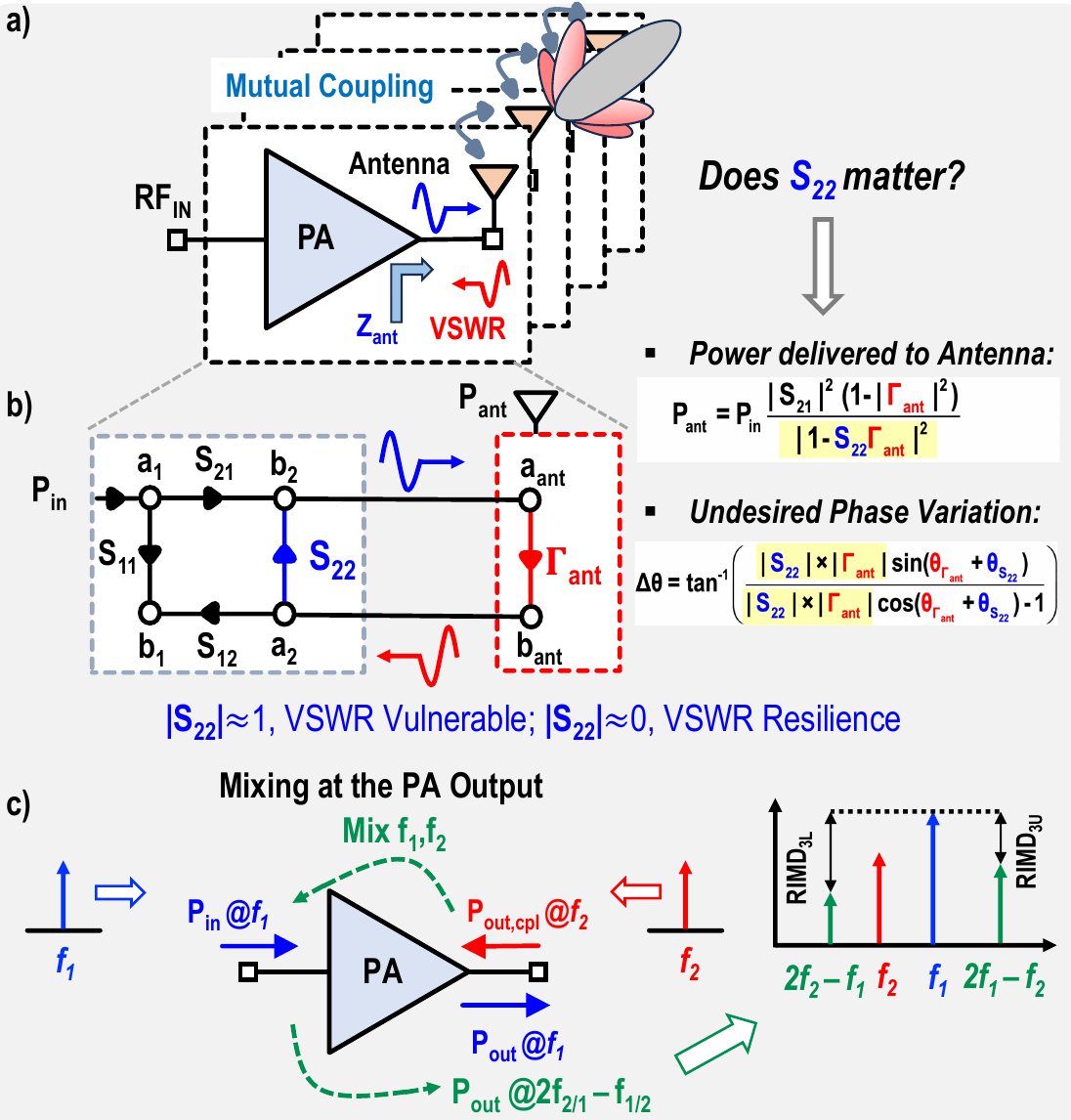}
    \caption{Antenna--PA interactions. (a) Load mismatch and mutual coupling in a phased-array TX. (b) Signal-flow modeling and analysis showing how PA output matching \(S_{22}\) affects delivered power and transmitted phase. (c) RIMD generated under reverse-coupled excitation.}
    \label{Fig:IIB}
    \vspace{-1.0em}
\end{figure}

Fig.~\ref{fig:antenna_vswr_infinite_array} illustrates the active antenna array VSWR using an infinite, tightly coupled phased array over 20--40~GHz as an example. The active VSWR varies with both frequency and scan angle and increases rapidly as the beam scans away from broadside. The H-plane scan is more severe than the E-plane scan, with the worst-case VSWR reaching $\sim$9:1 near 35~GHz at \(\theta=80^\circ\). Although extreme scan conditions can produce much worse VSWR, a large portion of the practical scan-angle and frequency region remains within or near the 3:1 boundary. Hence, 3:1 VSWR provides a useful common benchmark for comparing VSWR-resilient PA designs, while 4:1 or higher VSWR conditions serve as more severe stress-test cases~\cite{liu_2739-ghz_2025}.

The element-level PA performance variations can propagate to TX- and link-level requirements. Variations in \(\mathrm{OP}_{1\mathrm{dB}}\) and gain flatness reduce the usable average EIRP and link margin because TX-level margins must account for the worst-case VSWR condition. These effects should be considered at both the aggregate-signal level and the per-channel or per-subband level. Regulatory limits on EIRP may be specified in terms of power spectral density (PSD) over a reference bandwidth, e.g., dBm/100~MHz. Under such limits, load-dependent gain flatness can cause an individual channel or subband to reach the applicable PSD limit before the aggregate signal reaches its intended EIRP, reducing the usable average EIRP~\cite{pashaeifar_thesis_2024}.

For signals with a high peak-to-average power ratio (PAPR), \(P_{\mathrm{out}}\) degradation increases the required power back-off (PBO), lowers the average PAE (\(\mathrm{PAE}_{\mathrm{avg}}\)), and increases thermal density and inter-element thermal coupling among densely spaced array elements. Load-dependent AM--AM/AM--PM behavior increases the PA contribution to the TX EVM budget and reduces the margin available to other TX blocks. Severe VSWR conditions can also increase device voltage and current stress and reduce reliability margin. Collectively, these effects complicate system calibration, lookup-table (LUT) generation, and digital predistortion (DPD), and may require PA overdesign to satisfy the required performance specifications over the intended bandwidth and scan range~\cite{wang_millimeter-wave_2021, pashaeifar_chain-weaver_2024, sadhu_28-ghz_2017, alhamed_multi-band_2022, fager_linearity_2019}.

Together, these requirements define the system-level trade-off among modulation accuracy, link budget, implementation cost, and power consumption. Modulation accuracy and link budget determine system feasibility, while cost and power consumption determine the practical scalability of the solution. Implementation cost includes silicon area, calibration and test complexity, packaging and cooling requirements, and associated test time. The power budget depends on PA efficiency and the dc power consumed by sensing, control, and other front-end circuits. Calibration and reconfiguration can recover PA and array performance at the expense of additional time, hardware, and power overhead, whereas better intrinsic VSWR resilience may trade silicon area or circuit complexity for reduced calibration and test effort. The system-level consequences of antenna VSWR variation extend beyond passive mismatch loss. The following subsection analyzes this antenna--PA interaction.

\subsection{Antenna--PA Interaction Under VSWR Variation}

The PA--antenna interface is modeled using the signal-flow graph shown in Fig.~\ref{Fig:IIB}. In this review, the antenna is described by its reflection coefficient \(\Gamma_{\mathrm{ant}}\) and the PA by its small-signal \(S\)-matrix~\cite{eleraky_compact_2026}. The PA output matching is characterized by its output reflection coefficient \(S_{22}\). While conventional silicon integrated PA design has usually prioritized output loadline matching, PA output matching \(S_{22}\) becomes an important design parameter when the PA directly interfaces with a VSWR-varying antenna load.

Under the assumptions of a matched source and local linearization, the power delivered to the antenna, \(P_{\mathrm{ant}}\), for an arbitrary antenna reflection coefficient \(\Gamma_{\mathrm{ant}}\) can be expressed as~\cite{eleraky_compact_2026}
\begin{equation}
P_{\mathrm{ant}}
=
P_{\mathrm{in}}
\frac{|S_{21}|^2\left(1-|\Gamma_{\mathrm{ant}}|^2\right)}
{\left|1-S_{22}\Gamma_{\mathrm{ant}}\right|^2}.
\label{eq:power_delivered_antenna}
\end{equation}
Equation~\eqref{eq:power_delivered_antenna} separates the passive accepted-power factor, \(1-|\Gamma_{\mathrm{ant}}|^2\), from the PA-side reinteraction term, \(|1-S_{22}\Gamma_{\mathrm{ant}}|^{-2}\). Under the ideal matched-antenna condition, \(\Gamma_{\mathrm{ant}}=0\), and the influence of \(S_{22}\) disappears from this relation. This is the reason why conventional PA design methodologies often ignore \(|S_{22}|\) as long as a perfect \(50~\Omega\) load is assumed. However, under load mismatch (VSWR event), for which \(\Gamma_{\mathrm{ant}}\neq0\), the delivered power \(P_{\mathrm{ant}}\) becomes sensitive to the complex product \(S_{22}\Gamma_{\mathrm{ant}}\) in the denominator. Designing the PA with a low output reflection coefficient, i.e., small \(|S_{22}|\), can reduce the PA-side contribution to delivered power variation under antenna VSWR.

Beyond delivered power degradation, antenna VSWR variation also introduces transmitted phase variation.
Following the phase decomposition in~\cite{eleraky_compact_2026}, the total transmitted output phase under antenna load variation can be expressed as
\begin{equation}
\begin{aligned}
\text{Total Phase}
={}&
\phi_{21}
+
\underbrace{
\tan^{-1}\left(
\frac{
|\Gamma_{\mathrm{ant}}|\sin\phi_{\mathrm{ant}}
}{
1+|\Gamma_{\mathrm{ant}}|\cos\phi_{\mathrm{ant}}
}
\right)
}_{\text{(a) Load variation}}
\\
&-
\underbrace{
\tan^{-1}\left(
\frac{
-|S_{22}||\Gamma_{\mathrm{ant}}|
\sin(\phi_{S22}+\phi_{\mathrm{ant}})
}{
1-|S_{22}||\Gamma_{\mathrm{ant}}|
\cos(\phi_{S22}+\phi_{\mathrm{ant}})
}
\right)
}_{\text{(b) PA output mismatch}},
\end{aligned}
\label{eq:phase_total}
\end{equation}
where \(\Gamma_{\mathrm{ant}}=|\Gamma_{\mathrm{ant}}|e^{j\phi_{\mathrm{ant}}}\), \(S_{22}=|S_{22}|e^{j\phi_{S22}}\), and \(\phi_{21}\) is the nominal forward-transmission phase. Equation~\eqref{eq:phase_total} decomposes the transmitted phase error into two distinct components: 1) one arising from VSWR load variation and 2) the other from the PA output mismatch. Designing for low \(|S_{22}|\) can also minimize the transmitted phase error component from the PA output mismatch. For a given VSWR magnitude, the remaining phase error depends only on the antenna load phase and can be detected using a load impedance sensor and corrected through the beamforming coefficients. This enables a one-dimensional (1-D) LUT that maps the antenna load phase to the required phase
correction~\cite{eleraky_compact_2026,munzer_broadband_2023,liu_2739-ghz_2025}.

Beyond delivered power and transmitted phase variations, antenna mutual coupling can also introduce reverse intermodulation distortion (RIMD). As illustrated in Fig.~\ref{Fig:IIB}(c), signals from nearby TXs can couple into the PA output, mix with the PA carrier, and produce unwanted intermodulation products that degrade spectral-emission performance~\cite{atanasov_reverse_2020, ghorbanpoor_332_2026,eleraky_204_2026, eleraky_compact_2026}. A low output reflection coefficient, i.e., low \(|S_{22}|\), can help mitigate RIMD by reducing reverse excitation at the PA output. Since the PA output matching can vary with the input-power level, maintaining low \(|S_{22}|\) from small-signal operation to large-signal saturation is particularly important.

\subsection{Small-Signal and Large-Signal Evaluation}

Section~II-B provides the theoretical background on how \(S_{22}\) affects PA sensitivity to antenna VSWR variation. Practical PA operation, however, involves large RF swings and power-dependent output impedance. Therefore, VSWR resilience must be validated over the intended large-signal operating region. Large-signal scattering-parameter (LSSP) measurements can further indicate whether the output reflection coefficient \(S_{22}\) remains low as \(P_{\mathrm{out}}\) increases.

Continuous-wave (CW) large-signal measurements provide \(\mathrm{OP}_{1\mathrm{dB}}\), \(P_{\mathrm{sat}}\), PAE, and AM--AM/AM--PM behavior at nominal load. When connected to a load tuner or antenna-derived load set, the same large-signal measurement flow evaluates how the PA behavior changes over the relevant VSWR magnitude and load-angle region.

The transmitted waveform determines where this large-signal validation should be emphasized. Low-PAPR or near-constant-envelope signals place the PA close to the peak-power region, where robustness around \(\mathrm{OP}_{1\mathrm{dB}}\) or \(P_{\mathrm{sat}}\) is most relevant. By contrast, high-PAPR QAM, OFDM, and carrier-aggregation (CA) signals place the average \(P_{\mathrm{out}}\) (\(P_{\mathrm{avg}}\)) several decibels below the peak \(P_{\mathrm{out}}\), often by 6--12~dB depending on the waveform and linearity requirement. VSWR resilience must also be examined at PBO. This operating regime is increasingly important not only for wideband communication links, but also for future radar and sensing systems that may employ complex waveforms and concurrent multibeam operation. Under VSWR variation, \(\mathrm{PAE}_{\mathrm{avg}}\), \(P_{\mathrm{avg}}\), and EVM jointly characterize the performance and robustness of the PA~\cite{wang_millimeter-wave_2021,pashaeifar_chain-weaver_2024,liu_24_2025,eleraky_compact_2026}.

Nevertheless, in practice, to reduce the impact of active antenna VSWR in large scale high density arrays, a rule of thumb target is to ensure a PA small-signal \(S_{22}\) of approximately \(-8\) to \(-10~\mathrm{dB}\), or lower, while it is often observed that the PA large-signal \(S_{22}\) shows better matching because the PA power devices operate more in the triode or compression region under large signal drive.

Across these regimes, the reported VSWR resilience depends on the load region, excitation condition, and whether the PA performance is maintained under a given operating state or recovered after reconfiguration. The next design question is how the large-signal loadline matching that sets PA performance can be made compatible with the low-\(|S_{22}|\) condition that reduces VSWR sensitivity.

\begin{figure}[t!]
    \centering
    \includegraphics[width=0.9\linewidth]{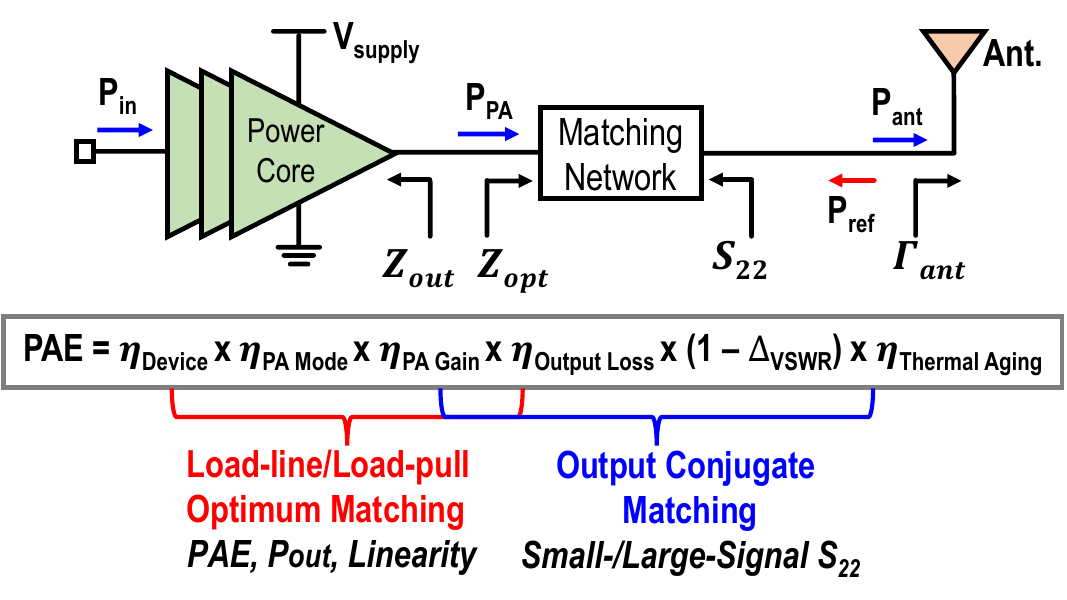}
    \caption{Conceptual PA power flow and PAE factors under antenna VSWR.}
    \label{fig:loadpull_output_match_concept}
    \vspace{-1.5em}
\end{figure}

\subsection{PA Output Conjugate- and Loadline-Matching Co-Design}

Classical PA design usually focuses on the large-signal load condition. The output matching network (OMN) transforms the nominal load \(R_{\mathrm{L}}\), typically \(50~\Omega\), to the optimum load-line/load-pull impedance \(Z_{\mathrm{opt}}\) seen by the PA core. This design flow is effective when the antenna or measurement load is fixed or tightly controlled.

For VSWR-resilient PAs, Section~II-B highlights the importance of PA output matching \(S_{22}\). Since delivered power degradation and transmitted phase error depend on the complex product \(S_{22}\Gamma_{\mathrm{ant}}\), a low output reflection coefficient can reduce PA sensitivity to antenna VSWR variation. In conventional large-signal PA design, however, PA output conjugate matching is often not treated as a primary objective. A common design intuition is that output conjugate matching and loadline matching are difficult to meet simultaneously. This comes from the distinction between the optimum loadline impedance \(Z_{\mathrm{opt}}\) and the PA output impedance \(Z_{\mathrm{out}}\), as shown in Fig.~\ref{fig:loadpull_output_match_concept}.

For instance, an ideal high-output-resistance transconductor with \(r_o \gg R_{\mathrm{L}}\) can support efficient power delivery to the transformed load while retaining a high intrinsic output resistance, which generally results in poor small-signal output matching, i.e., a large \(\lvert S_{22}\rvert\). One straightforward way to force output conjugate matching is to add a resistive termination or rely on dissipative OMN loss, but this directly introduces an efficiency penalty~\cite{pecile_study_2025}. From the small-signal maximum-power-transfer perspective, output conjugate matching can be interpreted, in the purely resistive case, as dissipating comparable power in \(R_{\mathrm{L}}\) and the resistive component of \(Z_{\mathrm{out}}\). Yet, this interpretation is most applicable to weak-signal or Class-A operation. Pecile \textit{et al.} show that the conjugate-matching efficiency penalty decreases when the PA operates in Class-AB, Class-B, and beyond, and can be reduced through appropriate drive and bias conditions~\cite{pecile_study_2025}.

In this view, low \(|S_{22}|\) forms part of the PA core design used to reduce PA sensitivity to antenna VSWR variation while preserving the intended large-signal operation. In the remainder of this review, this design principle is referred to as \emph{simultaneous output and loadline matching (SOLM)}, analogous in design philosophy to simultaneous noise and input matching (SNIM) in low-noise amplifier (LNA) designs~\cite{eleraky_55_2025,lin_v-e-band_2026}.

\section{Circuit and Architecture Techniques for VSWR-Resilient PAs}

This section reviews representative circuit and architecture techniques for mitigating PA sensitivity to antenna VSWR variation. Following the antenna--PA interaction framework in Section~II, the discussion emphasizes the operating principle, implementation requirements, and suitability of each technique for dense mm-Wave and cm-Wave phased arrays.

\begin{figure}[tp]
    \centering
    \includegraphics[width=0.65\linewidth]{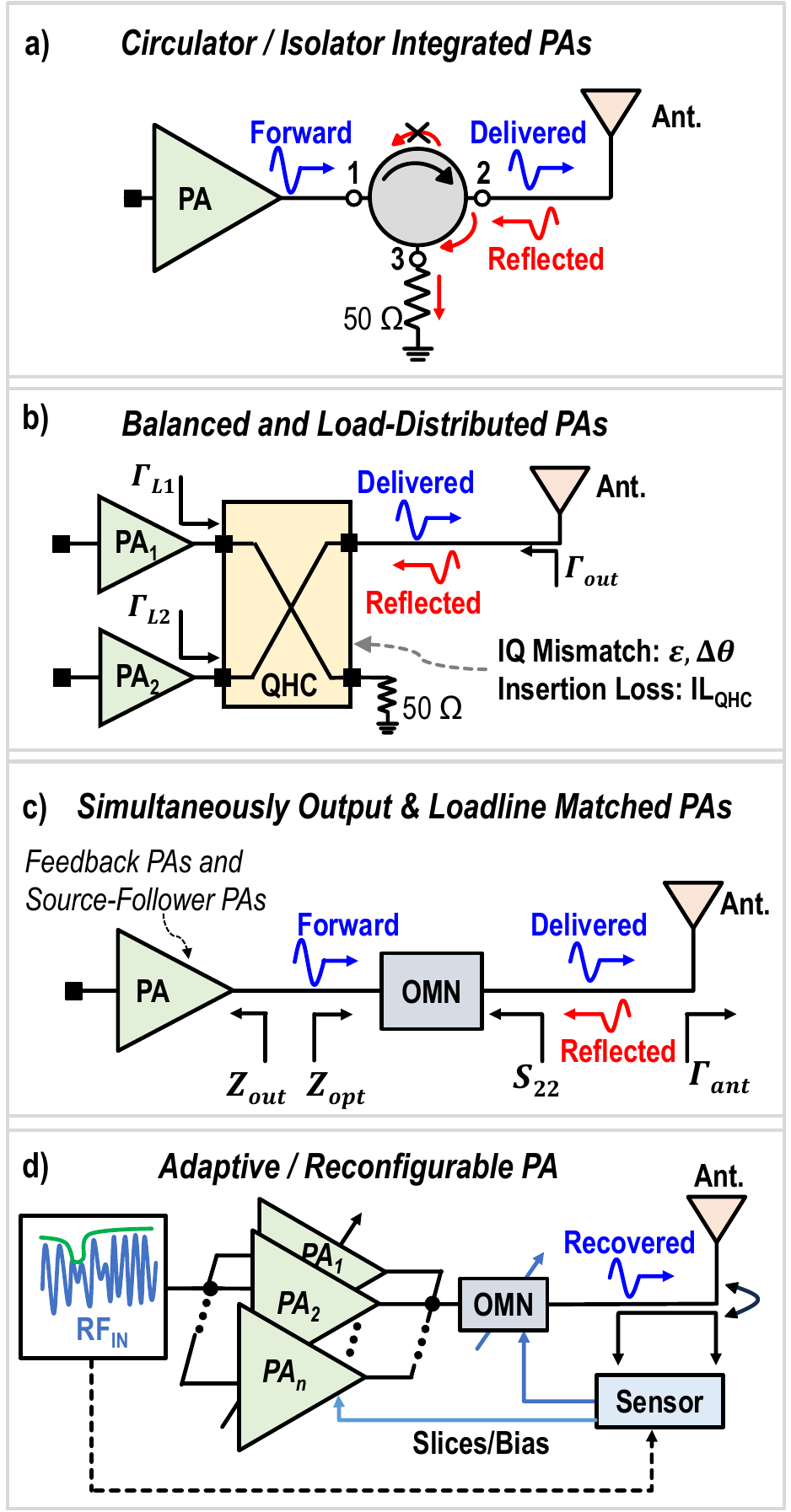}
    \caption{Representative circuit and architecture approaches for reducing PA sensitivity to antenna VSWR variation: (a) circulator/isolator-integrated PAs; (b) low-\(|S_{22}|\) output matching by passive networks in balanced and load-distributed PAs; (c) low-\(|S_{22}|\) output matching by inherent active PA core designs; and (d) sensing-assisted adaptive/reconfigurable PAs.}
    \label{fig:vswr_pas}
    \vspace{-1.0em}
\end{figure}

\subsection{Circulator- and Isolator-Based PA Interfaces}

A common solution employs nonreciprocal ferrite-based isolators to decouple the PA from the antenna by absorbing reflected signals. Ideally, an isolator allows the PA output signal to propagate toward the antenna while absorbing the signal reflected by the antenna. A circulator provides a related three-port function, where the signal generated by the TX is routed to the antenna and the signal incident from the antenna is routed to another port. As shown in Fig.~\ref{fig:vswr_pas}(a), circulator- or isolator-based PA interfaces can reduce the impact of antenna impedance variation on the PA load condition and protect the PA from antenna back-reflections~\cite{nagulu_non-magnetic_2021,nagulu_non-reciprocal_2020}.

Ferrite-based circulators and isolators are widely used in RF and microwave systems, but their magnetic biasing, bulky form factor, insertion loss, and operating-frequency constraints hinder their use in dense mm-Wave and cm-Wave phased arrays~\cite{eleraky_compact_2026,nagulu_non-magnetic_2021}. This limitation has motivated magnet-free nonreciprocal components that are compatible with semiconductor integration. Representative approaches include active-transistor-based, N-path-filter-based, time-modulated, and spatio-temporally modulated networks~\cite{reiskarimian_magnetic-free_2016,reiskarimian_182_2017,reiskarimian_analysis_2018,dinc_synchronized_2017,nagulu_non-reciprocal_2020}. For example, spatio-temporal conductivity modulation has demonstrated magnet-free passive non-reciprocity in a fully integrated 25-GHz CMOS circulator, showing a possible path toward integrated mm-Wave circulator/isolator interfaces~\cite{dinc_millimeter-wave_2017,dinc_synchronized_2017}.

For PA applications, the nonreciprocal interface must be evaluated together with the PA and the antenna. Important considerations include TX-to-antenna insertion loss, reverse isolation, operating bandwidth, power handling, silicon area, clocking or modulation overhead, and large-signal behavior. Recent PA-integrated isolator/circulator demonstrations show the potential of this direction for VSWR-resilient PA operation, while also highlighting that the achieved resilience can depend on the large-signal operating condition and the output power regime~\cite{pashaeifar_millimeter-wave_2026,ghorbanpoor_332_2026}. Therefore, circulator- and isolator-based PA interfaces are attractive when the benefit of PA/load decoupling outweighs the added passive loss, area overhead, and implementation complexity.

\subsection{Output Matching by Passive Networks: Balanced and Load-Distributed PAs}

Balanced and load-distributed PAs mitigate antenna VSWR variation by redistributing the antenna-reflected wave among multiple PA paths, which prevents the full reflected signal from perturbing a single PA output. The PA active devices are not output matched, while the output matching is achieved using various balanced passive networks instead of passive loss. In a conventional balanced PA (BPA), quadrature couplers redirect reflected power to an isolation port terminated with a resistive load, as illustrated in Fig.~\ref{fig:vswr_pas}(b). This reduces the dependence of the combined PA response on the VSWR angle~\cite{pashaeifar_millimeter-wave_2021,pashaeifar_144_2021,quaglia_effect_2022,quaglia_mitigation_2022,nikandish_unbalanced_2021,diverrez_24-31ghz_2023,diverrez_22-44_2024}.

While BPAs are often regarded as \emph{VSWR-free}, this holds only under idealized assumptions. The common argument is that the two branch PAs see equal-and-opposite load mismatches, so their nonlinear degradations are equal-and-opposite and cancel at the combiner. In practice, this is only approximately true. For a finite branch PA output matching $S_{22}$, the loads presented to the two PAs by an antenna reflection \(\Gamma_{\mathrm{ant}}\) are
\begin{equation}
\label{eq:BPA_Loads}
\Gamma_{L1}
=
\frac{\Gamma_{\mathrm{ant}}}
{1+S_{22}\Gamma_{\mathrm{ant}}},
\qquad
\Gamma_{L2}
=
\frac{-\Gamma_{\mathrm{ant}}}
{1-S_{22}\Gamma_{\mathrm{ant}}},
\end{equation}
where $\Gamma_{L1}$ and $\Gamma_{L2}$ are opposite in sign but \emph{unequal in magnitude}, becoming exactly opposite only in the ideal limit \(S_{22}\to0\)~\cite{pashaeifar_millimeter-wave_2021}. The distortion, power, and gain variations of the two branches are not exactly opposite and do not completely cancel, leaving residual distortion and variation at the antenna port. The approximately opposing branch loads lead to a balanced operation in which one amplifier sees a lower effective impedance, with higher \(P_{\mathrm{out}}\) but lower PAE, while the other sees a higher effective impedance, with lower \(P_{\mathrm{out}}\) but higher PAE. This provides partial mismatch compensation after recombination, but it does not eliminate large-signal deviation, load-dependent branch behavior, or device stress~\cite{pashaeifar_millimeter-wave_2021,pashaeifar_144_2021}.

Multi-way load-distribution architectures extend the same principle by spreading the antenna-reflected signal over more PA paths. For example, chain-weaver BPAs use staged quadrature combining and phase rotation to provide multiple effective loads distributed over the VSWR circle, improving gain, phase, and linearity consistency under VSWR variation~\cite{pashaeifar_chain-weaver_2024,pashaeifar_327_2024}.

In practice, the in-phase/quadrature (I/Q) mismatch of the quadrature hybrid coupler (QHC) can introduce additional combining power loss, power leakage into the termination resistors, and imperfect output matching, causing gain variations~\cite{pashaeifar_chain-weaver_2024}. For the BPA illustrated in Fig.~\ref{fig:vswr_pas}(b), assuming QHC amplitude and phase errors of $\epsilon$ and $\Delta\theta$, respectively, the output reflection coefficient \(\Gamma_{\mathrm{out}}\) can be calculated as
\begin{equation}
\label{eq:Gamma_out_IQMM}
\begin{aligned}
\left|\Gamma_{\mathrm{out}}\right|
={}&
\frac{\sqrt{1+(1+\epsilon)^4-2(1+\epsilon)^2\cos(2\Delta\theta)}}
{1+(1+\epsilon)^2}
\\[-0.2em]
&\quad\times
\left|\Gamma_{\mathrm{PA}}\right| \cdot
\left|\mathrm{IL}_{\mathrm{QHC}}\right|^2,
\end{aligned}
\end{equation}
where \(\mathrm{IL}_{\mathrm{QHC}}\) is the insertion loss of the QHC~\cite{pashaeifar_chain-weaver_2024}.

In summary, the main limitations are quadrature imbalance, coupler loss, isolation-port dissipation, combining-network footprint, bandwidth, and thermal handling of redistributed reflected power. Thus, balanced and load-distributed PAs are attractive when reduced load-angle dependence outweighs the passive loss and area penalty of the combining network.

\subsection{Output Matching by Inherent Active PA Core Designs: Feedback PAs and Source-Follower PAs}

An alternative direction is to make VSWR resilience an inherent property of the PA and its output network. In this approach, the active PA core is designed for inherently low output reflection or low effective output impedance. Therefore, a complex balanced passive network is not required solely to obtain a matched external port. Output-matched PAs reduce sensitivity to antenna VSWR through a small output reflection coefficient \(|S_{22}|\) or a low effective output impedance, without relying primarily on external isolation, reflected-wave redistribution, or load-dependent state selection~\cite{eleraky_204_2026,eleraky_55_2025,eleraky_compact_2026,pecile_study_2025}.

As discussed in Section~II, small \(|S_{22}|\) reduces the PA-side \(S_{22}\Gamma_{\mathrm{ant}}\)-dependent variations in delivered power and transmitted phase. It can also attenuate reverse excitation and help reduce RIMD, although the nonlinear reverse-excitation response should be validated separately. Yet, achieving low \(|S_{22}|\) without compromising \(P_{\mathrm{out}}\) and efficiency is challenging in practical silicon PAs~\cite{wang_246_2020,pecile_study_2025,eleraky_compact_2026}.

Therefore, the key design goal is to achieve high-efficiency PA active device cores with inherent output matching (Fig.~\ref{fig:vswr_pas}(c)), i.e., simultaneous output matching and loadline matching (SOLM), analogous to simultaneous noise and input matching (SNIM) in LNA design~\cite{lin_v-e-band_2026, eleraky_compact_2026}.

\begin{figure}[tp]
    \centering
    \includegraphics[width=0.7\linewidth]{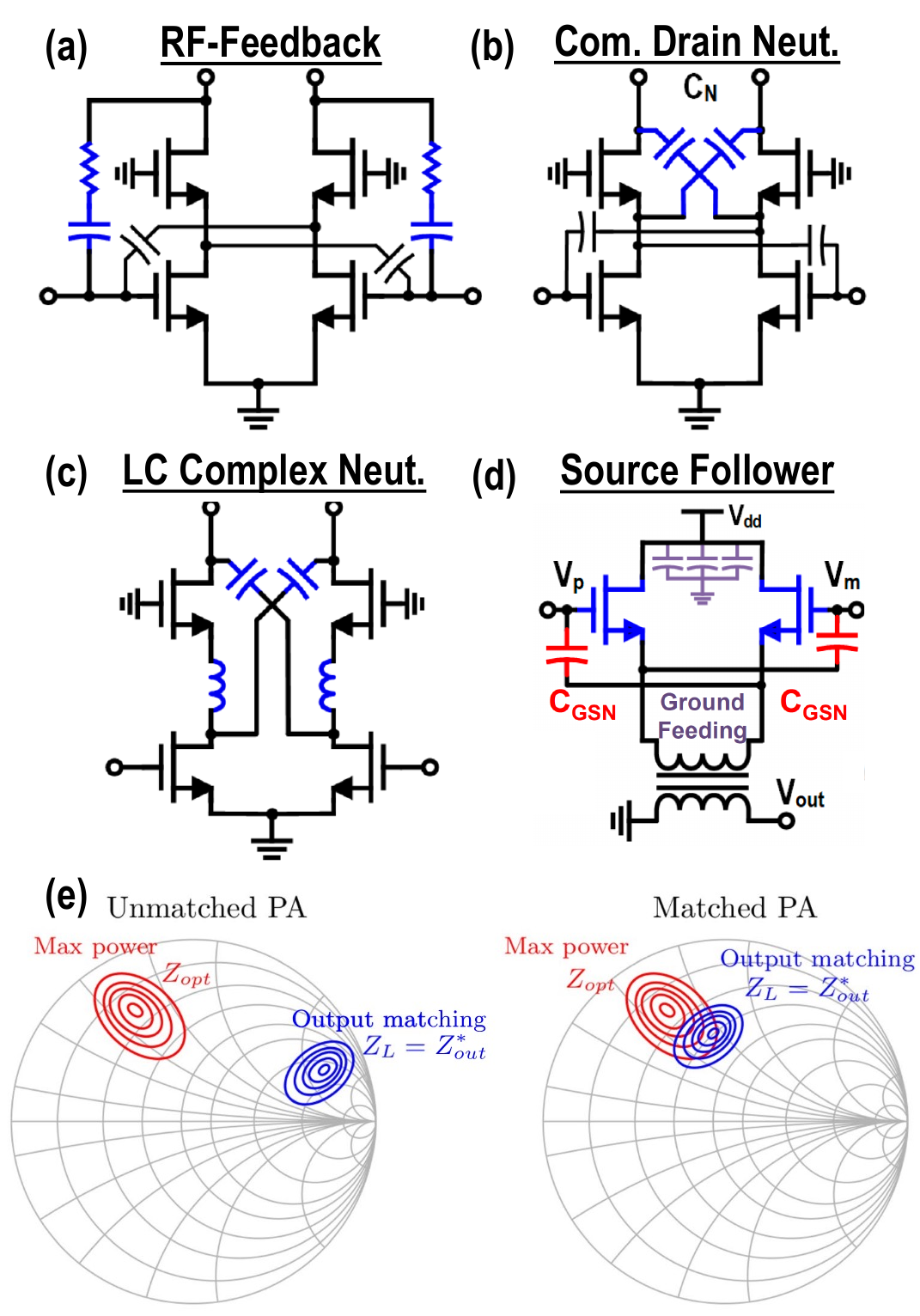}
    \caption{PA active core circuits with low \(S_{22}\). (a) RC feedback. (b) Drain--source neutralization. (c) Complex-cascode LC neutralization. (d) Source-follower output stage~\cite{aref_26_2015,eleraky_204_2026}. (e) Loadline and output matching contours for conventional unmatched PAs and output-matched (SOLM) PAs~\cite{pecile_study_2025}.}
    \label{fig:output_matched}
    \vspace{-1.5em}
\end{figure}

Fig.~\ref{fig:output_matched} summarizes representative circuit routes for improving \(S_{22}\) or lowering the effective PA output impedance. Traditionally, feedback architectures have been used to improve output matching of RF PAs at GHz operating frequencies. RC negative feedback can reduce output impedance at cm-Wave and mm-Wave frequencies, but the lossy feedback (Fig.~\ref{fig:output_matched}(a)) path may introduce stability concerns and degrade large-signal performance. Drain--source neutralization (Fig.~\ref{fig:output_matched}(b)) can also reduce output impedance, but a larger neutralization capacitance increases the output capacitance and output quality factor, which can elevate matching-network loss. Complex-LC feedback/neutralization (Fig.~\ref{fig:output_matched}(c)) provides additional design freedom to align the output-matching condition with the desired loadline matching~\cite{eleraky_55_2025,eleraky_compact_2026,eleraky_204_2026}. Source-follower (common-drain) output stages provide another direct route to low output impedance and improved \(S_{22}\) (Fig.~\ref{fig:output_matched}(d)). The source follower can provide current gain and power gain, while its voltage gain remains at or below unity. It requires a large RF input-voltage swing to generate the desired output swing, increasing the drive and inter-stage matching requirements of the preceding stage~\cite{aref_26_2015,eleraky_204_2026}.

\subsection{Adaptive and Reconfigurable PAs}

Adaptive and reconfigurable PAs mitigate antenna VSWR variation by restoring a favorable PA loadline/load-pull condition after the load state is known, estimated, or calibrated. Instead of relying on a single fixed output impedance transformation, these techniques employ reconfigurable PA cores, tunable/reconfigurable matching networks, or both to adjust the PA operating condition in response to impedance variations, as illustrated in Fig.~\ref{fig:vswr_pas}(d). The objective is to recover selected large-signal and modulation metrics, such as \(P_{\mathrm{sat}}\), \(\mathrm{OP}_{1\mathrm{dB}}\), PAE, gain flatness, AM--AM/AM--PM, or EVM, over the intended VSWR magnitude and load-angle region~\cite{singh_inverted_2023,singh_pa_2023,liu_24_2025,guo_1-d_2023,chappidi_multi-port_2020,mannem_reconfigurable_2020,bowers_integrated_2013}.

Reconfigurable matching networks (RMNs) provide a direct example of this approach. By changing the impedance transformation between the PA and the antenna, an RMN can map different antenna impedances closer to the desired large-signal load-line/load-pull region. Recent compact RMN demonstrations use switched-capacitor banks and coupled-line matching sections to extend the reconfiguration range, while integrated power sensors can support gain estimation and power monitoring for VSWR-aware operation~\cite{liu_24_2025,liu_2739-ghz_2025,munzer_broadband_2023}. Such approaches are particularly useful when the VSWR load region is bounded and can be characterized by beam state, frequency, or embedded sensing.

Adaptive operation can also be implemented through PA-core or load-modulated architectures~\cite{wang_super-resolution_2019,wang_artificial-intelligence_2019,mannem_reconfigurable_2020,mannem_broadband_2021,xu_-field_2020}. For example, reconfigurable Doherty PAs can adjust the relative gain, phase, or operating mode of the Main/Auxiliary amplifiers to compensate antenna impedance variation and recover large-signal performance under VSWR~\cite{hu_antenna_2015,mannem_reconfigurable_2020}. These Doherty and load-modulated examples are treated as adaptive performance-recovery techniques, since their VSWR resilience relies on changing the load-modulation or control state according to the antenna VSWR variation condition.

\subsection{Sensing-Assisted Monitoring and Control}

Power, gain, and impedance sensing circuits provide the information required for VSWR-aware PA operation. In phased arrays, PA large-signal performance can degrade due to device aging, thermal effects, and antenna VSWR variation. Since VSWR-related power degradation can occur over a much shorter time scale than long-term aging, compact VSWR-resilient sensing circuits are useful for detecting \(P_{\mathrm{out}}\) degradation, monitoring PA gain, estimating the load condition, and supporting array calibration or reconfiguration~\cite{munzer_broadband_2023,liu_2739-ghz_2025,liu_3210_2024}.

The sensed quantity must be chosen carefully. Voltage-only or current-only sensing can track \(P_{\mathrm{out}}\) accurately only when the load impedance is known. Under an unknown complex load, true-power sensing or impedance-aware sensing is more directly related to the real power delivered to the antenna. Gain sensing is also useful because antenna VSWR variation changes both the delivered power and the PA gain profile, which affects AM--AM flatness, EIRP regulation, beamforming calibration, and EVM margin~\cite{munzer_single-ended_2021,munzer_broadband_2022,munzer_single-ended_2022,munzer_broadband_2023,liu_2739-ghz_2025}.

Sensing circuits do not by themselves improve \(|S_{22}|\) or recover the optimum PA loadline. Instead, they enable VSWR-aware operation when the sensed information is used for built-in self-test (BiST), output-power regulation, LUT generation, state selection, beamforming calibration, or closed-loop reconfiguration. Relevant metrics include sensing error and dynamic range, reference-plane accuracy, frequency coverage, RF loading, dc power, and silicon area. In large-scale arrays, these overheads are important because sensing and calibration complexity scale with the number of elements~\cite{pashaeifar_chain-weaver_2024,liu_2739-ghz_2025}.

These techniques mitigate antenna VSWR variation through different mechanisms, while their reported results are not directly interchangeable. A fair comparison should state the RF reference plane and VSWR load region, the excitation and control state, the maintained or recovered metric, and the associated implementation overhead.

\begin{figure}[t!]
    \centering
    \includegraphics[width=0.9\linewidth]{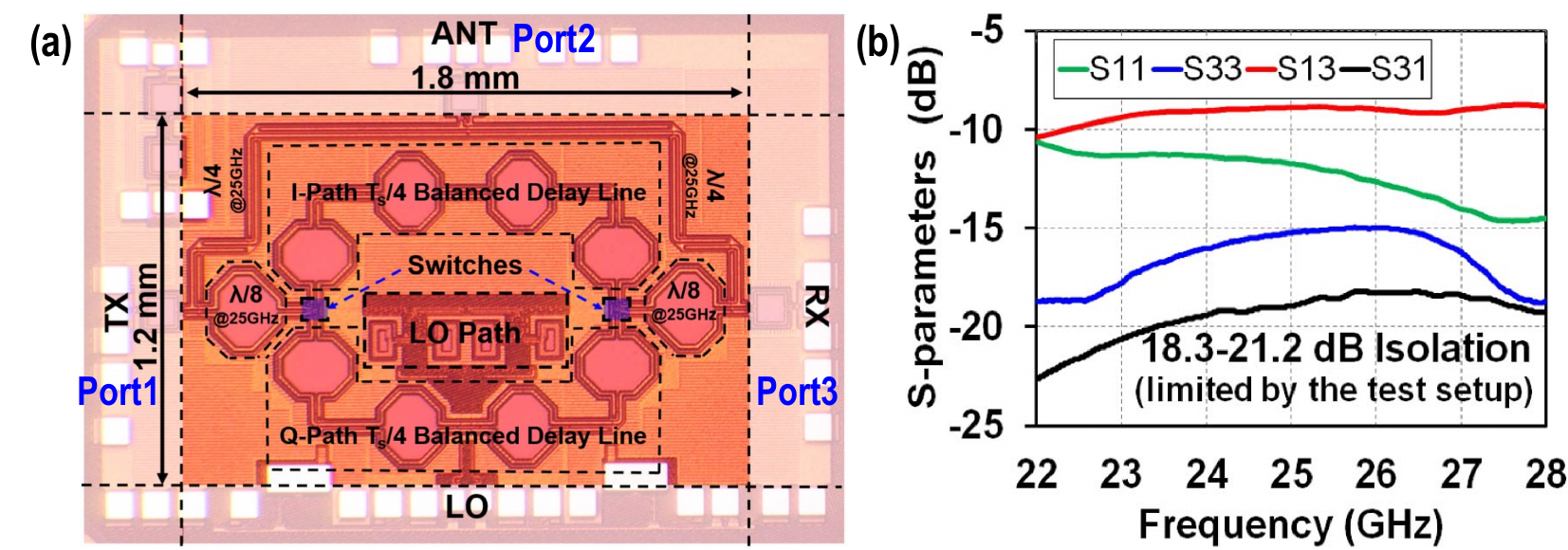}
    \caption{A 25-GHz nonreciprocal, magnet-free passive circulator in 45-nm SOI CMOS. (a) Chip micrograph. (b) S-parameter measurement results~\cite{dinc_millimeter-wave_2017}.}
    \label{fig:on_chip_circulator}
    \vspace{-0.5em}
\end{figure}

\begin{figure}[t]
    \centering
    \includegraphics[width=0.9\linewidth]{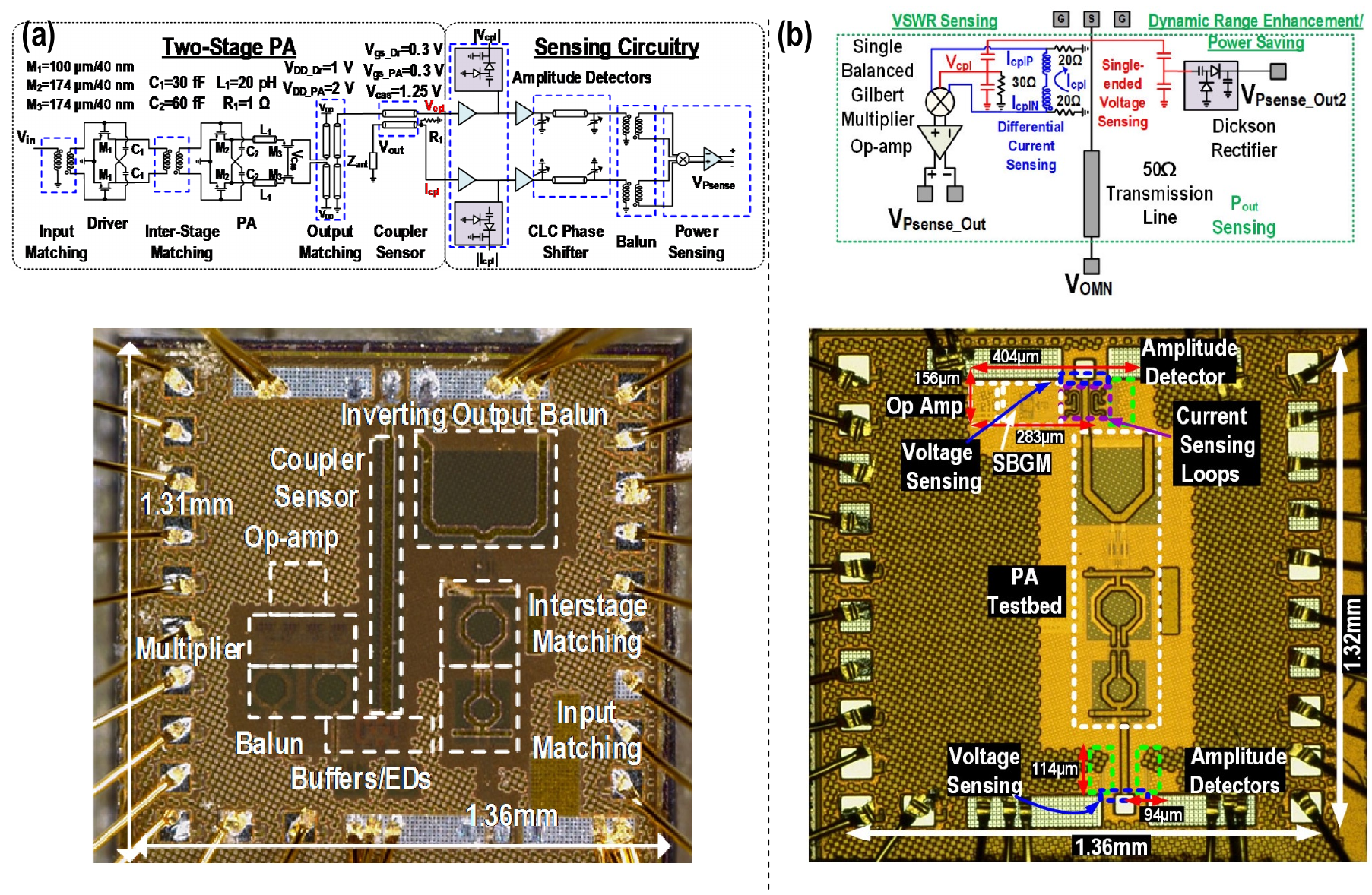}
    \caption{Sensing circuits for VSWR events. (a) A single-ended quadrature-coupler-based joint true-power/impedance sensor in 45-nm SOI~\cite{munzer_single-ended_2022}. (b) A compact 27--39-GHz true-power-and-gain sensor in 45-nm SOI~\cite{liu_3210_2024,liu_2739-ghz_2025}.}
    \label{fig:Sensors_PA}
    \vspace{-1.0em}
\end{figure}

\begin{figure}[t]
    \centering
    \includegraphics[width=0.85\linewidth]{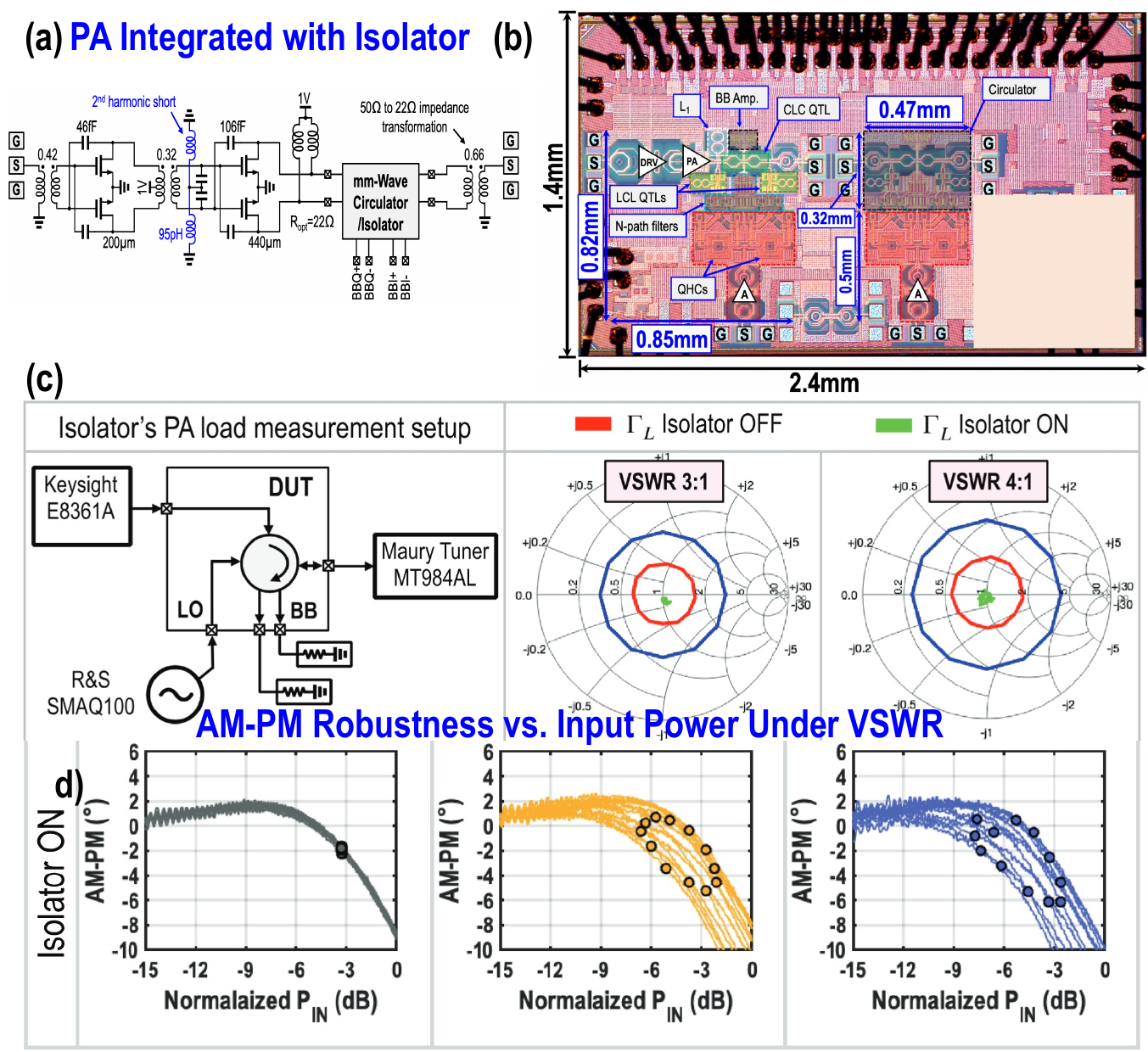}
    \caption{A PA with an integrated N-path-filter-based isolator in 40-nm CMOS~\cite{pashaeifar_millimeter-wave_2026}. (a) Circuit diagram. (b) Chip micrograph. Measured (c) load presented by the isolator and (d) AM--PM vs. \(\mathrm{P}_{\mathrm{in}}\) under 3:1 and 4:1 VSWR.}
    \label{fig:PA_integrated_isolator}
    \vspace{-0.5em}
\end{figure}

\section{Recent VSWR-Resilient PA Examples}

This section presents illustrative mm-Wave and cm-Wave circuit demonstrations of the PA architectures reviewed above. Some architectures maintain VSWR-resilient performance through nonreciprocal isolation, passive reflected-wave redistribution, or inherent PA core output matching. Adaptive architectures instead recover performance through load-dependent reconfiguration. The reported resilience is quantified through large-signal variation over VSWR angle and frequency, together with \(S_{22}\), silicon area, and other implementation overheads. A full-angle 3:1 VSWR circle is used as the primary demonstration condition~\cite{wang_millimeter-wave_2021,mannem_reconfigurable_2020,hu_antenna_2015,liu_24_2025,liu_2739-ghz_2025,munzer_broadband_2022}.

\subsection{Circulator- and Isolator-Based Circuits}
Circulator- and isolator-based circuits use a nonreciprocal interface to decouple the PA from antenna impedance variation. Ideally, the PA output signal is delivered toward the antenna, while the antenna-reflected signal is routed to another port or absorbed by a termination.

Fig.~\ref{fig:on_chip_circulator}(a) and (b) show a 25-GHz SOI CMOS circulator that uses spatiotemporal conductivity modulation to provide magnet-free passive nonreciprocity at mm-Wave frequencies~\cite{dinc_millimeter-wave_2017,dinc_synchronized_2017}. The circuit achieves minimum TX-to-ANT (\(S_{21}\)) and ANT-to-RX (\(S_{32}\)) insertion losses (ILs) of 3.3 and 3.2~dB, respectively, with a 4.6-GHz 1-dB bandwidth. The TX-to-RX isolation (\(S_{31}\)) is \(-18.3\) to \(-21.2\)~dB, limited by the measurement setup. The TX-to-ANT and ANT-to-RX input-referred \(\mathrm{P}_{1\mathrm{dB}}\) values are \mbox{\(>+21.5\)} and \(>+21\)~dBm, respectively. The circulator consumes 78.4~mW and has a core chip area of \(2.16~\mathrm{mm}^{2}\). This work demonstrates the feasibility of on-chip nonreciprocal circuits, while the area overhead and additional power consumption remain the main constraints on their viability in phased-array systems.

\begin{figure}[t]
    \centering
    \includegraphics[width=0.85\linewidth]{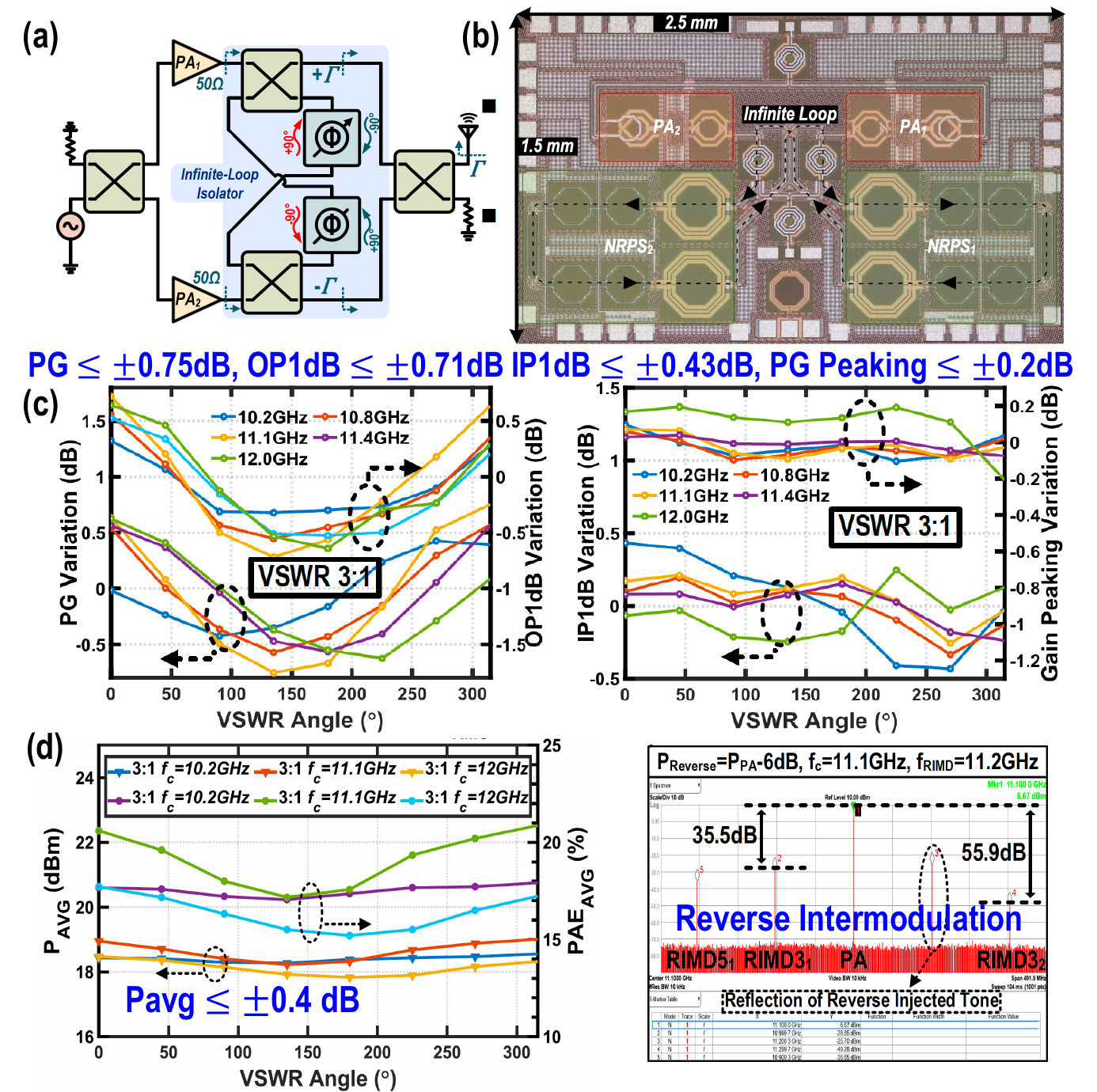}
    \caption{An infinite-loop isolator-enabled balanced PA in 22-nm FD-SOI~\cite{ghorbanpoor_332_2026}. (a) Circuit diagram. (b) Chip micrograph. Measured results under 3:1 VSWR (c) PG, \(\mathrm{OP}_{1\mathrm{dB}}\), \(\mathrm{IP}_{1\mathrm{dB}}\), and PG peaking. (d) \(P_{\mathrm{avg}}\), \(\mathrm{PAE}_{\mathrm{avg}}\), and RIMD.}
    \label{fig:infinite_loop}
    \vspace{-1.0em}
\end{figure}

\begin{figure}[t]
    \centering
    \includegraphics[width=0.8\linewidth]{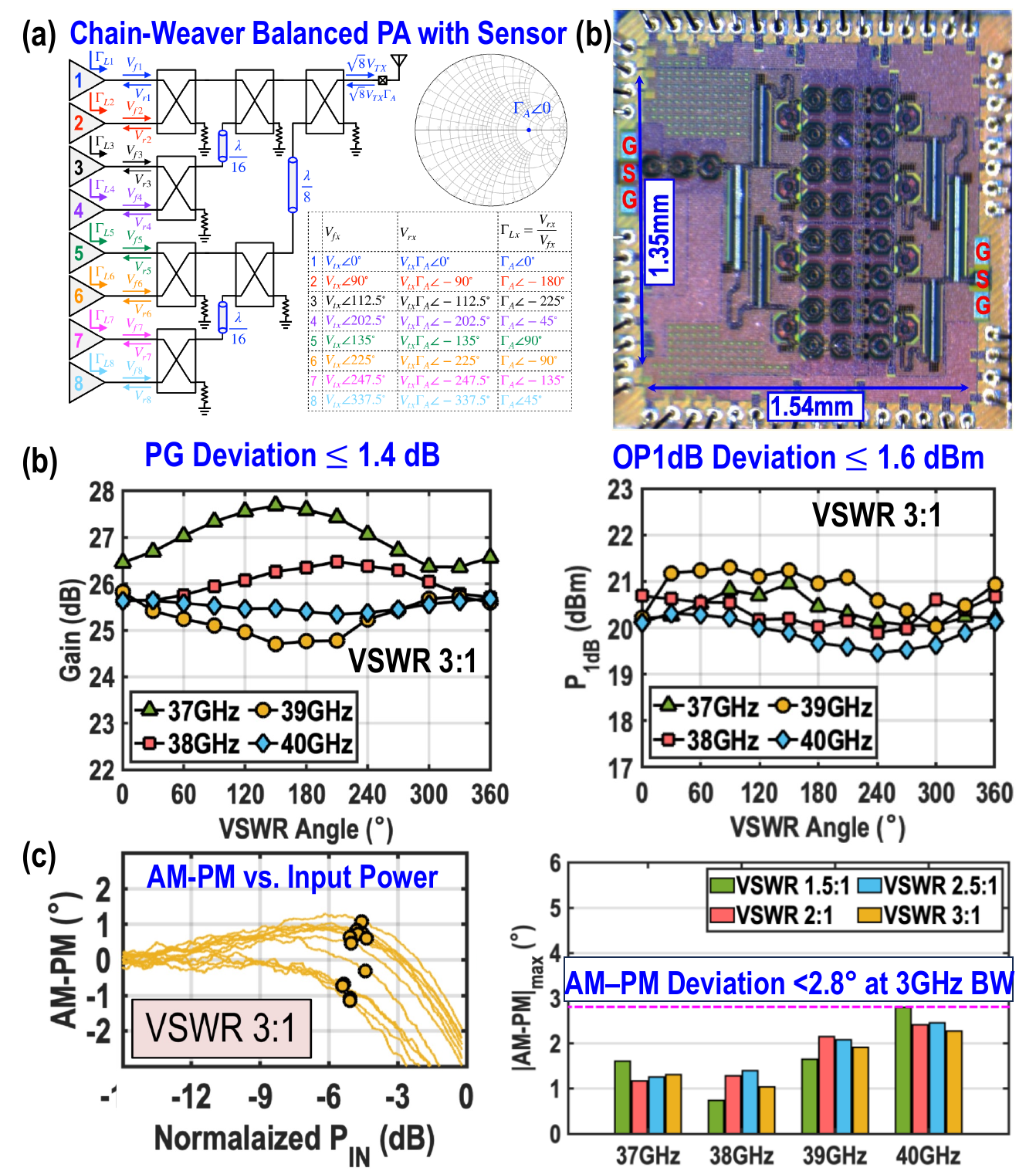}
    \caption{An eight-way chain-weaver balanced PA with embedded impedance/power sensor in 40-nm CMOS~\cite{pashaeifar_chain-weaver_2024}. (a) Circuit diagram. (b) Chip micrograph. Measured (c) PG, \(\mathrm{OP}_{1\mathrm{dB}}\), and AM--PM under 3:1 VSWR.}
    \label{fig:chain_weaver}
    \vspace{-1.0em}
\end{figure}

In addition to nonreciprocal load decoupling, accurate VSWR-aware sensing can support PA interfaces that rely on BiST, monitoring, or reconfiguration for performance recovery. A single-ended quadrature-coupler-based joint true-power/impedance sensor in 45-nm SOI CMOS demonstrates this role using an integrated PA testbed at 38~GHz (Fig.~\ref{fig:Sensors_PA}(a)). It senses the real power delivered to a complex antenna load under 3:1 VSWR with a power sensing error (PSE) within \(\pm3.35\)~dB, while also estimating the antenna impedance with maximum \(|\Gamma|\) and \(\angle\Gamma\) errors of 0.238 and \(28.9^{\circ}\), respectively. It achieves a 16-dB dynamic range at \(50~\Omega\), a sensor area of \(0.456~\mathrm{mm}^{2}\), and a power consumption of 44.5~mW~\cite{munzer_single-ended_2022}. A 27--39-GHz compact true-power-and-gain sensor further improves this direction by reducing the sensor area to 0.044~\(\mathrm{mm}^{2}\) and the dc power consumption to 24.8~mW, while maintaining PSEs within \(\pm1.5\) and \(\pm3\)~dB and dynamic ranges above 18.69 and 17.65~dB under 2:1 and 3:1 VSWR, respectively~\cite{liu_3210_2024,liu_2739-ghz_2025} (Fig.~\ref{fig:Sensors_PA}(b)). These works mainly demonstrate power, impedance, and gain sensing under VSWR. PSE and dynamic range are key performance metrics, while dc power consumption and sensor area quantify implementation overhead when VSWR-resilient sensing is used to support PA correction or recovery in large-scale arrays.

Fig.~\ref{fig:PA_integrated_isolator}(a) and (b) show a 27.1--31.1-GHz PA with an integrated isolator in 40-nm bulk CMOS~\cite{pashaeifar_millimeter-wave_2026}. The isolator is reconfigured from a magnet-free N-path-filter-based CMOS circulator and occupies an ultra-compact core area of 0.38~\(\mathrm{mm}^{2}\). The stand-alone isolator shows \(>15\)-dB reverse isolation. The PA achieves a measured 15.15-dBm peak \(P_{\mathrm{out}}\) and 33\% drain efficiency (DE). When the isolator power consumption is included, the corresponding peak PAE is 15.1\%. The complete PA-isolator front end occupies a core area of 0.7~\(\mathrm{mm}^{2}\).

Fig.~\ref{fig:PA_integrated_isolator}(c) shows the measurement setup and results. The red/green lines are the loads seen by the PA when the isolator is OFF/ON, respectively. The green line remains confined near the optimum load even when the antenna VSWR is increased to 4:1, confirming that the isolator can effectively decouple the PA load from the antenna mismatch. The measured AM--PM results show relatively high VSWR resilience at lower \(P_{\mathrm{out}}\). When the PA approaches peak \(P_{\mathrm{out}}\), the AM--PM improvement degrades due to the saturation of the switches in the isolator. This result highlights both the benefit and the high-power limitation of the integrated isolator-based PA, along with its power consumption and overall efficiency penalties.

\subsection{Balanced and Load-Distributed PAs}

Balanced and load-distributed PA examples reduce PA/TX sensitivity to VSWR variation by redistributing the reflected wave through quadrature or multi-way combining networks.

The chain-weaver eight-way balanced PA in Fig.~\ref{fig:chain_weaver}(b) extends the conventional balanced PA concept to a multi-path mm-Wave PA~\cite{pashaeifar_chain-weaver_2024,pashaeifar_327_2024}. In the event of antenna impedance mismatch, the chain-weaver combining network provides multiple effective loads distributed on the VSWR circle, so that the combined performance is averaged over multiple PA paths. The 40-nm CMOS implementation combines eight Class-AB PA paths, operates from 35 to 43~GHz, occupies a \(2.08~\mathrm{mm}^{2}\) core area, and achieves \(S_{22}<-20\)~dB, \(P_{\mathrm{sat}}=25.19\)~dBm, 16.19\% peak PAE, and \(\mathrm{OP}_{1\mathrm{dB}}>22\)~dBm over 36--42~GHz.

With modulated signals, the chain-weaver PA supports a 2-GHz 64-QAM OFDM signal at 16-dBm \(P_{\mathrm{avg}}\) with about \(-25\)-dB EVM and an 800-MHz 256-QAM OFDM signal at 12.17-dBm \(P_{\mathrm{avg}}\) with EVM below \(-30.3\)~dB without DPD. Over a full-angle 3:1 VSWR circle, the BPA exhibits a PG variation of \(\pm0.7\)~dB, an \(\mathrm{OP}_{1\mathrm{dB}}\) variation of \(\pm0.8\)~dB, and an AM--PM variation below \(2.8^{\circ}\) over a 3-GHz bandwidth. The embedded impedance/power sensor further supports output power regulation, BiST, and load-based performance optimization. At 3:1 VSWR, the reported impedance-angle and magnitude errors are $12.3^\circ$ and 0.106, respectively~\cite{pashaeifar_chain-weaver_2024}.

\begin{figure}[t]
    \centering
    \includegraphics[width=0.98\linewidth]{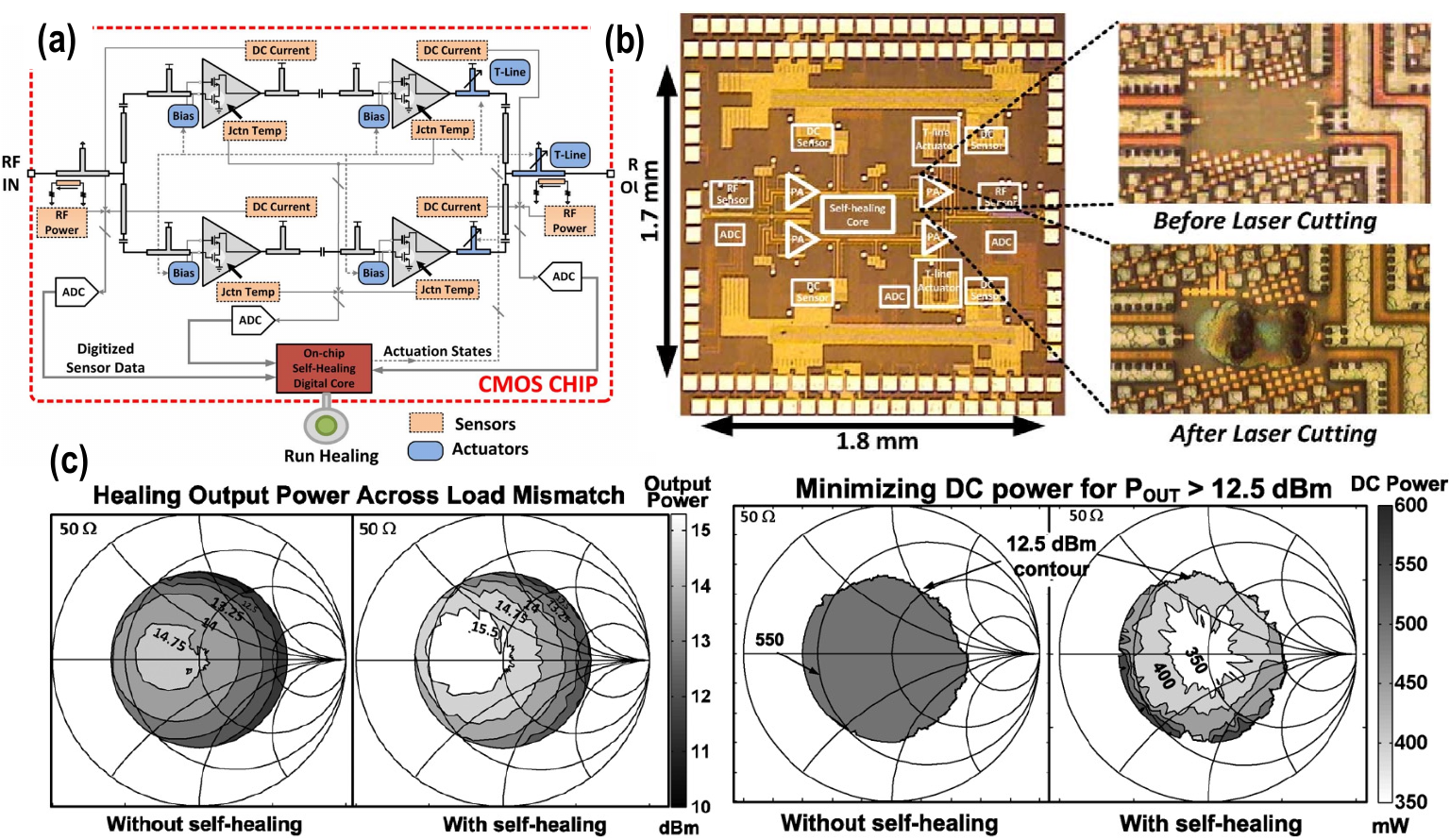}
    \caption{A 28-GHz integrated self-healing PA~\cite{bowers_integrated_2013}. (a) Circuit diagram. (b) Chip micrograph. (c) Contours before and after self-healing for maximum \(P_{\mathrm{out}}\) and the dc power at \(P_{\mathrm{out}}\) of 12.5 dBm over the entire 4:1 VSWR circle.}
    \label{fig:Self-Healing}
    \vspace{-1.0em}
\end{figure}

The infinite-loop isolator-enabled PA in Fig.~\ref{fig:infinite_loop} provides another integrated nonreciprocal route~\cite{ghorbanpoor_332_2026}. The architecture uses a CMOS-compatible infinite-loop isolator to maintain the intended PA load condition while attenuating reflected signals. The 22-nm FD-SOI implementation operates around 10.2--12~GHz and reports \(S_{22}<-15\)~dB, \(P_{\mathrm{sat}}=23.9\)~dBm, \(\mathrm{OP}_{1\mathrm{dB}}=23.3\)~dBm, and peak PAE of 30.6\% excluding clock power and 28\% including clock power. Under 3:1 VSWR, the reported gain, \(\mathrm{OP}_{1\mathrm{dB}}\), and average-power variations remain within 1~dB. Under the stated reverse-excitation condition, \(\mathrm{RIMD}_{3}\) remains below approximately \(-35.5\)~dBc, further connecting isolator-enabled VSWR resilience with suppression of reverse-coupled excitation.

\subsection{Adaptive and Reconfigurable PAs}

Adaptive and reconfigurable PAs recover large-signal performance by changing the PA operating state or output impedance transformation according to the antenna VSWR. The following examples quantify how closely measured PA performance can be restored after load-dependent state selection.

An early fully integrated adaptive example is the 28-GHz CMOS self-healing PA in~\cite{bowers_integrated_2013}, as shown in Fig.~\ref{fig:Self-Healing}. Its on-chip RF-power, dc-current, and temperature sensors, data converters, digital optimizer, bias controls, and tunable transmission-line stubs form a complete sensor-to-controller-to-actuator loop. Over calibrated loads within a 4:1 VSWR circle, the self-healing state expands the load region supporting high output power; a second optimization mode reduces dc power by up to 35\% while maintaining 12.5-dBm output power. For a 100-ksymbol/s 16-QAM signal at 12.5-dBm output power, the average EVM across ten chips improves from 5.9\% to 4.2\%. The reported exhaustive-search time is approximately 0.8~s, demonstrating complete on-chip integration while leaving faster beam-state adaptation as an open challenge.

\begin{figure}[t]
    \centering
    \includegraphics[width=0.8\linewidth]{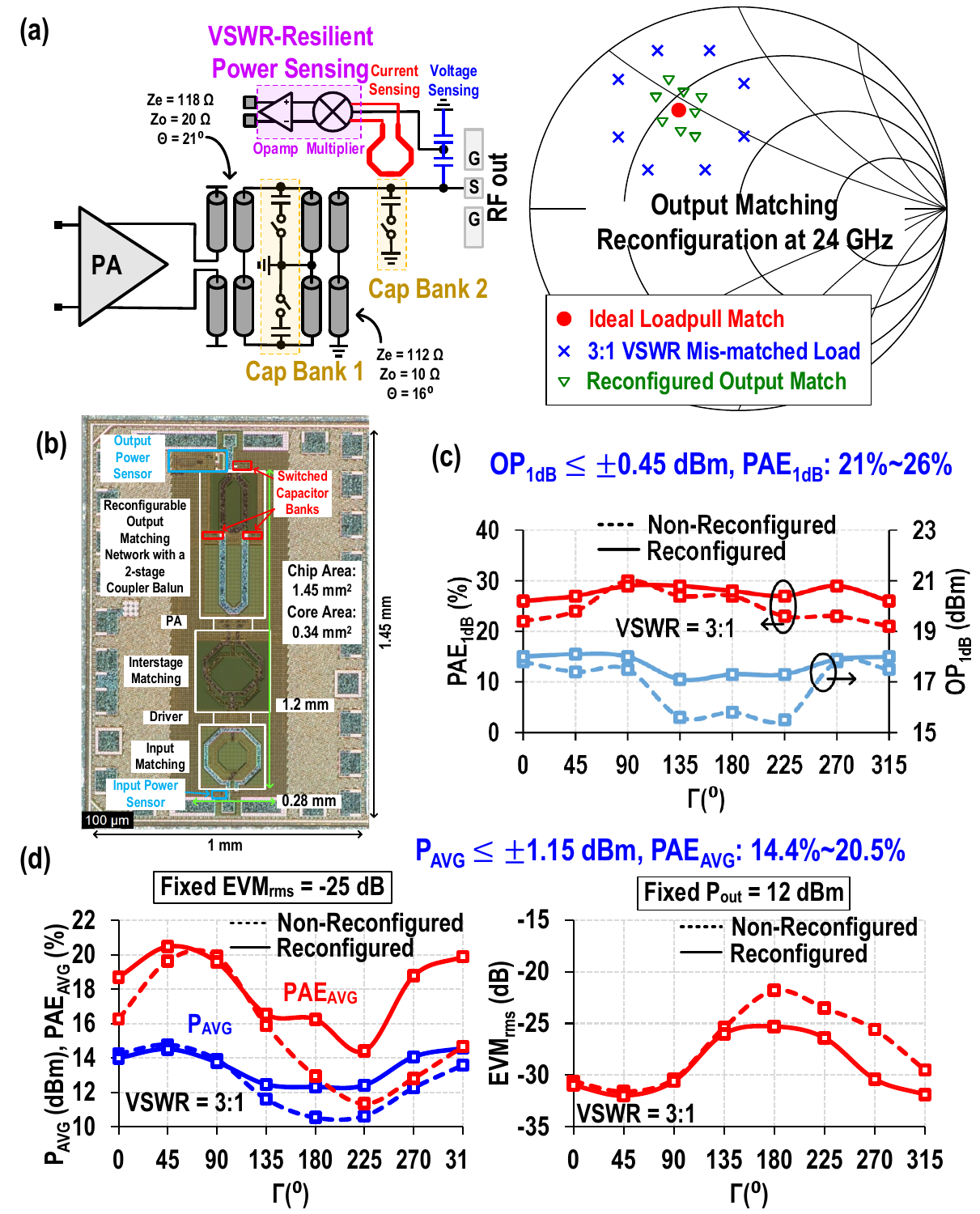}
    \caption{A compact 24-GHz RMN-based PA with integrated power sensors in 22-nm FDX+~\cite{liu_24_2025}. (a) Circuit diagram. (b) Chip micrograph. Measured (c) \(\mathrm{PAE}_{1\mathrm{dB}}\) and \(\mathrm{OP}_{1\mathrm{dB}}\). (d) \(P_{\mathrm{avg}}\), \(\mathrm{PAE}_{\mathrm{avg}}\), and \(\mathrm{EVM}_{\mathrm{rms}}\) under 3:1 VSWR.}
    \label{fig:RMN_PA}
    \vspace{-1.0em}
\end{figure}

The 24-GHz PA in GlobalFoundries 22-nm FDX+ shown in Fig.~\ref{fig:RMN_PA} demonstrates a compact reconfigurable output matching network (RMN) approach~\cite{liu_24_2025}. The RMN uses two coupled-line sections and two switched-capacitor banks to extend the load-impedance reconfiguration region and map a mismatched antenna load toward the desired PA load-line impedance. The PA occupies a 0.34-\(\mathrm{mm}^{2}\) core area and integrates input/output power sensors for gain estimation and power monitoring.

Under a nominal 50-\(\Omega\) load, the PA reports \(S_{11}<-10\)~dB from 18.8 to 25.5~GHz. At 24~GHz, it achieves 20-dB gain, \(\mathrm{OP}_{1\mathrm{dB}}=17.8\)~dBm, and \(\mathrm{PAE}_{1\mathrm{dB}}=26\%\). With a 100-MHz single-carrier 64-QAM signal at 24~GHz and a 50-\(\Omega\) load, it achieves \(P_{\mathrm{avg}}=13.2\)~dBm and \(\mathrm{PAE}_{\mathrm{avg}}=17\%\) at \(\mathrm{EVM}_{\mathrm{rms}}=-25.4\)~dB. Reconfiguration improves \(\mathrm{OP}_{1\mathrm{dB}}\) at every load angle on the full-angle 3:1 VSWR circle. Before reconfiguration, \(\mathrm{OP}_{1\mathrm{dB}}\) remains > 15.5~dBm with a 2.3-dB range. After reconfiguration, it is recovered to > 17.1~dBm with an approximately \(\pm0.45\)-dB variation, while the minimum \(\mathrm{PAE}_{1\mathrm{dB}}\) increases from > 21\% to > 26\%.

\begin{figure}[t!]
    \centering
    \includegraphics[width=0.85\linewidth]{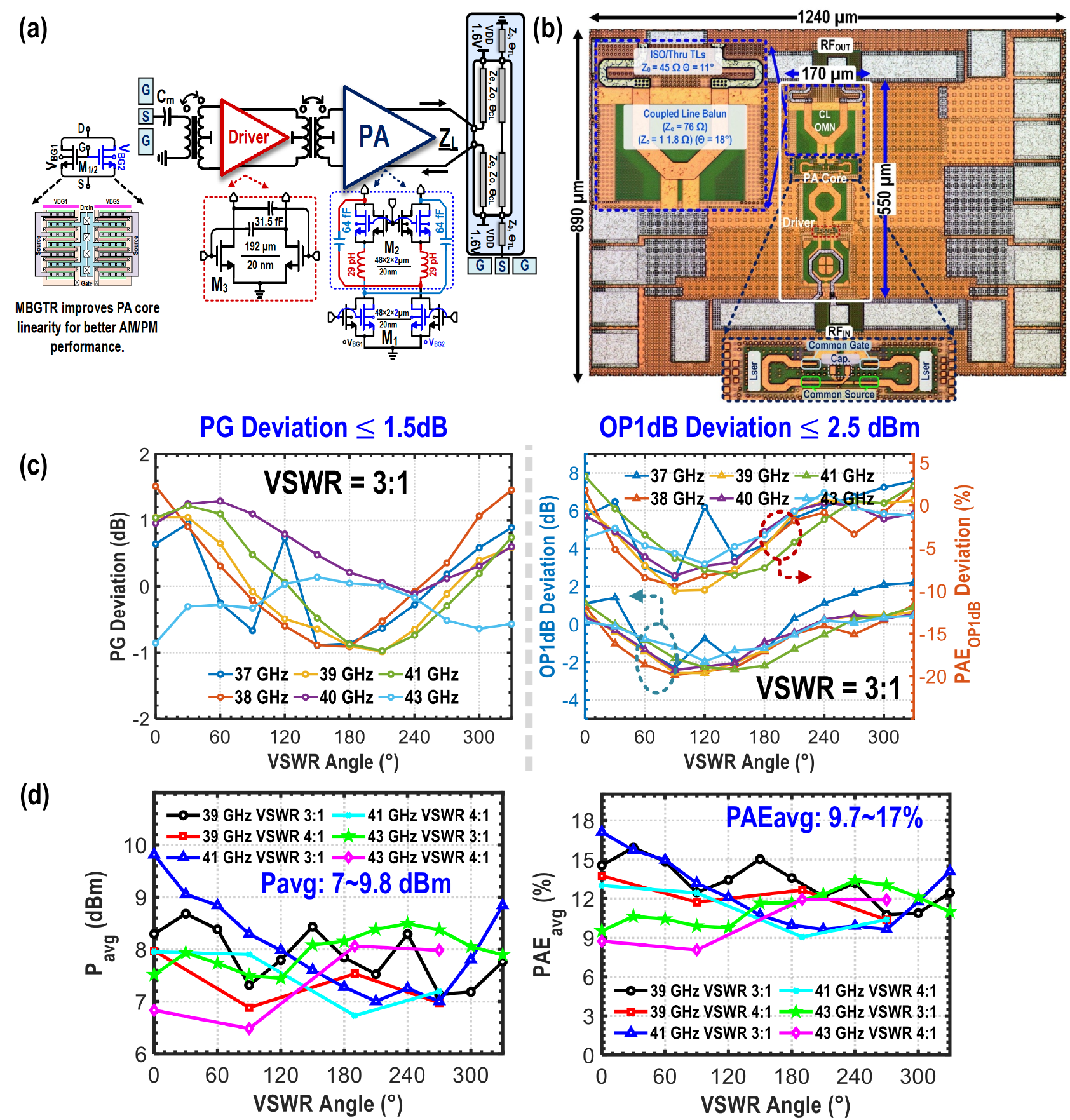}
    \caption{A complex-cascode LC-neutralized FR2 PA with low \(|S_{22}|\) in 22-nm FDX+~\cite{eleraky_compact_2026,eleraky_55_2025}. (a) Circuit diagram. (b) Chip micrograph. Measured results under 3:1 VSWR: (c) PG and \(\mathrm{OP}_{1\mathrm{dB}}\); (d) \(P_{\mathrm{avg}}\) and \(\mathrm{PAE}_{\mathrm{avg}}\).}
    \label{fig:complex_cascode_pa}
\end{figure}

For modulated-signal operation under the 3:1 VSWR circle, the PA is measured with a 100-MHz single-carrier 64-QAM signal. With the \(\mathrm{EVM}_{\mathrm{rms}}\) target fixed at \(-25\)~dB, the unreconfigured PA provides \(P_{\mathrm{avg}}>10.5\)~dBm and \(\mathrm{PAE}_{\mathrm{avg}}>11.3\%\). After reconfiguration, \(P_{\mathrm{avg}}\) and \(\mathrm{PAE}_{\mathrm{avg}}\) are recovered to > 12.3~dBm and > 14.4\%, respectively, with maximum reported \(P_{\mathrm{avg}}=14.6\)~dBm and maximum \(\mathrm{PAE}_{\mathrm{avg}}=20.5\%\). When \(P_{\mathrm{avg}}\) is fixed at 12~dBm, \(\mathrm{EVM}_{\mathrm{rms}}\) remains below \(-25\)~dB after reconfiguration, an improvement over the worst-case value of approximately \(-21\)~dB without reconfiguration.

\subsection{Output-Matched PAs for Inherent VSWR Resilience}

Output-matched PA demonstrations maintain VSWR-resilient performance by reducing the PA output reflection coefficient or the effective output impedance of the final power stage.

A complex-cascode LC-neutralized PA in Fig.~\ref{fig:complex_cascode_pa} demonstrates an output-matched approach in 22-nm FDX+ technology~\cite{eleraky_compact_2026,eleraky_55_2025}. The PA uses complex-LC neutralization in the cascode output stage to align the output conjugate-match condition with the optimum loadline impedance. The active core occupies approximately 0.093~\(\mathrm{mm}^{2}\). Under a nominal 50-\(\Omega\) load, it reports \(S_{22}<-10\)~dB from 32 to 62~GHz. At 39~GHz, it delivers \(P_{\mathrm{sat}}=15.7\)~dBm with 31.5\% peak PAE and \(\mathrm{OP}_{1\mathrm{dB}}=15.3\)~dBm with 31\% PAE. Across 35--45~GHz at a 1.6-V supply, \(P_{\mathrm{sat}}\) remains within 15.1--16.2~dBm, with peak PAE of 25--34.2\%.

Under a 3:1 VSWR circle with 30\(^{\circ}\) load-angle steps and a single bias setting~\cite{eleraky_compact_2026}, the worst-case power gain (PG) deviation is within 1.5~dB, the \(\mathrm{OP}_{1\mathrm{dB}}\) degradation is limited to 2.5~dB, and the \(\mathrm{PAE}_{1\mathrm{dB}}\) drop remains below 10 percentage points relative to the 50-\(\Omega\) case, from 37 to 43~GHz. Under the more severe 4:1 VSWR condition, the power-gain deviation remains within 1.5~dB, the \(\mathrm{OP}_{1\mathrm{dB}}\) degradation is below 3~dB, and the \(\mathrm{PAE}_{1\mathrm{dB}}\) drop remains below 11\%.

\begin{figure}[t!]
    \centering
    \includegraphics[width=0.85\linewidth]{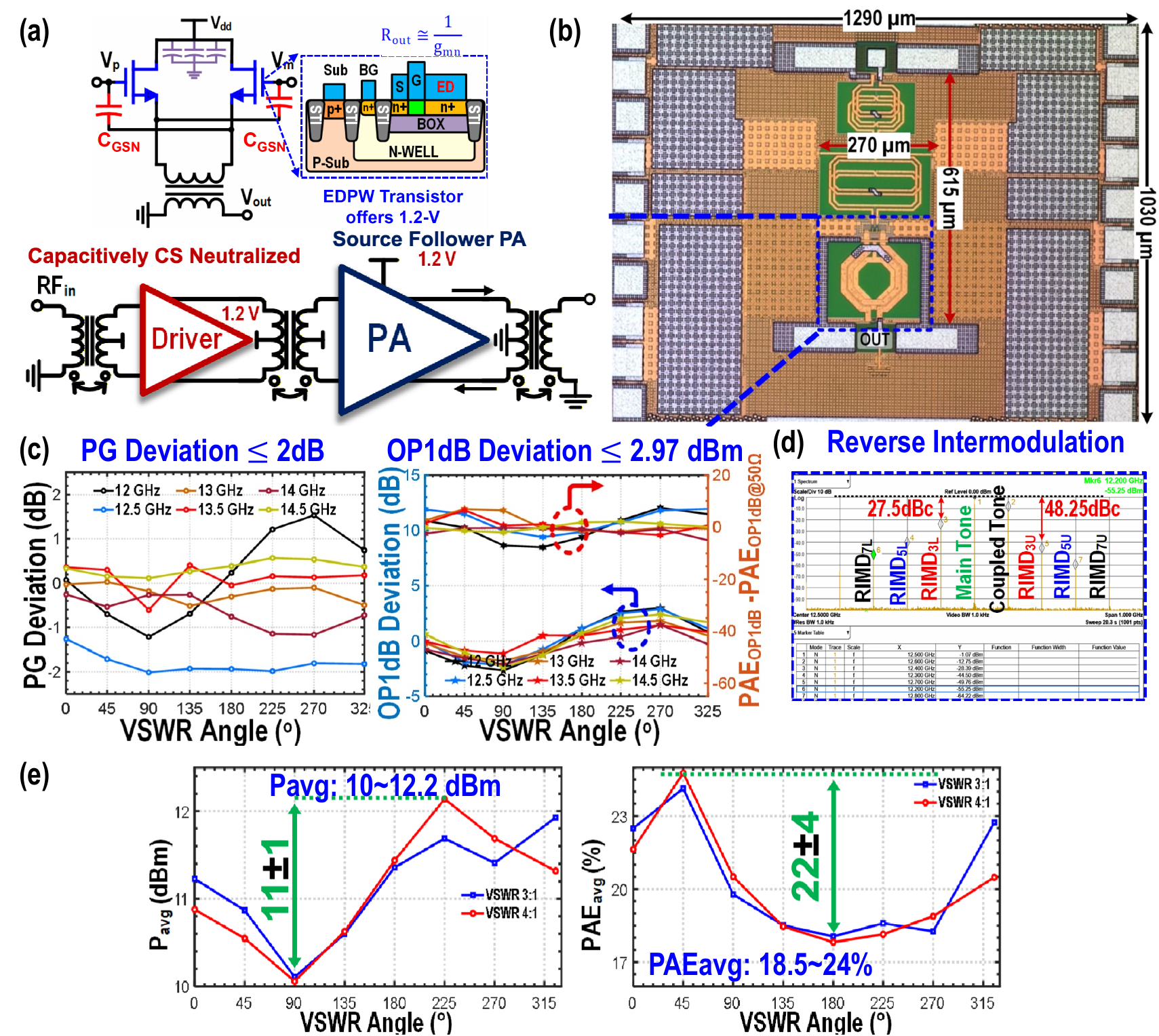}
    \caption{A source-follower FR3 output-matched PA in 22-nm FDX+~\cite{eleraky_204_2026}. (a) Circuit diagram. (b) Chip micrograph. Measured results under 3:1 VSWR (c) PG and \(\mathrm{OP}_{1\mathrm{dB}}\). (d) RIMD and (e) \(P_{\mathrm{avg}}\) and \(\mathrm{PAE}_{\mathrm{avg}}\).}
    \label{fig:source_follower_pa}
    \vspace{-0.5em}
\end{figure}

\begin{figure}[t]
    \centering
    \includegraphics[width=0.85\linewidth]{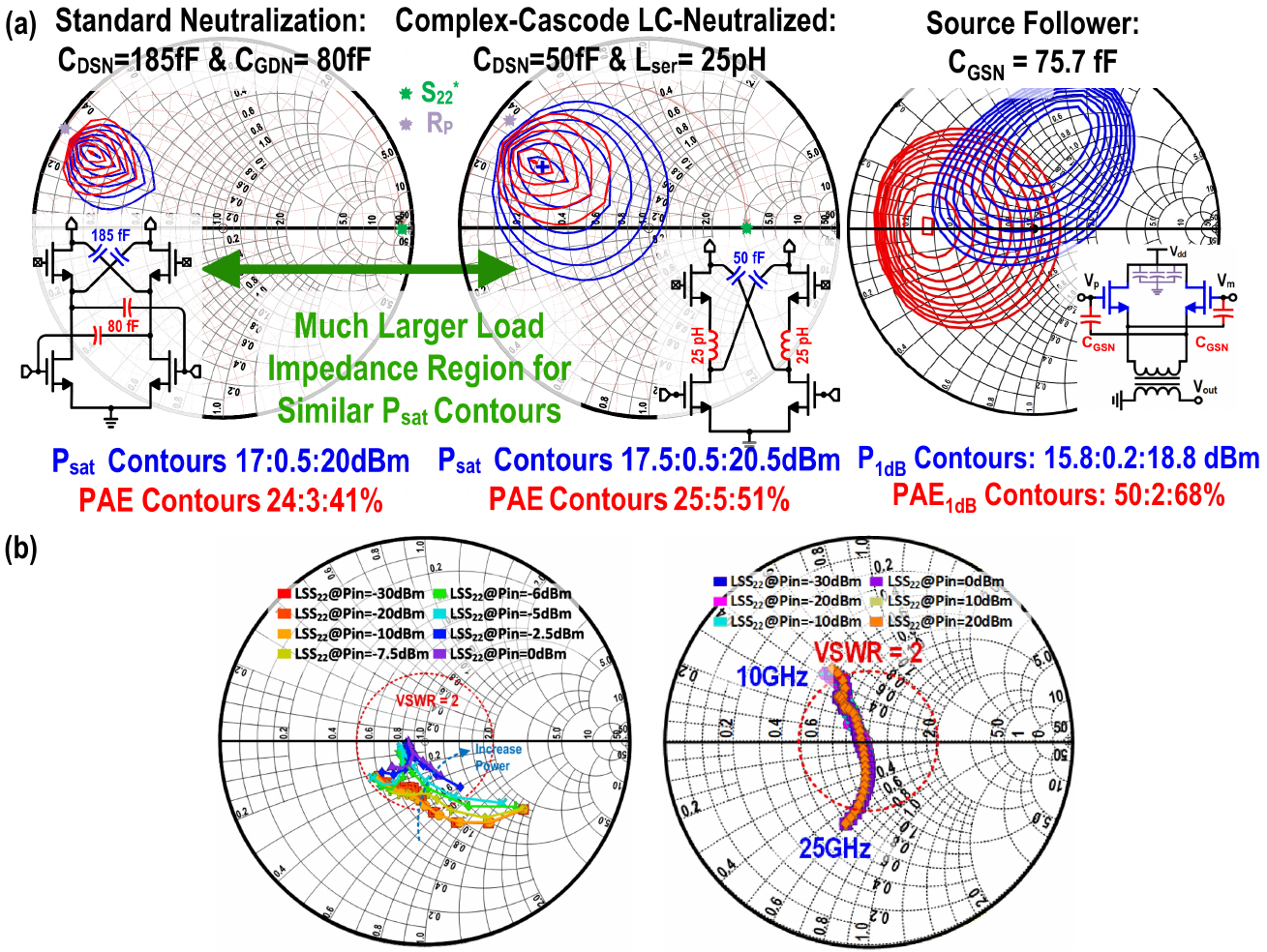}
    \caption{(a) \(P_{\mathrm{out}}\) and PAE large-signal load-pull contours for standard-neutralized cascode, complex-cascode LC-neutralized, and source-follower PA cores. (b) LSSP \(S_{22}\) trajectories versus \(P_{\mathrm{in}}\) over the respective operating frequency ranges of the LC-neutralized and source-follower PAs~\cite{eleraky_55_2025,eleraky_204_2026,eleraky_compact_2026}.}
    \label{fig:loadpull_output_match_comparison}
    \vspace{-1.0em}
\end{figure}

With a 100~MSym/s 64-QAM signal over a full-angle 3:1 VSWR circle, the PA reports \(P_{\mathrm{avg}}\) of 9.5, 9.1, and 8.5~dBm with \(\mathrm{PAE}_{\mathrm{avg}}\) of 14.8\%, 14.4\%, and 11.7\% at 39, 41, and 43~GHz, respectively. For VSWR up to 4:1 across 39--43~GHz, the measured \(\mathrm{EVM}_{\mathrm{rms}}\) remains below \(-24.5\)~dB, while \(P_{\mathrm{avg}}\) and \(\mathrm{PAE}_{\mathrm{avg}}\) exceed 6.8~dBm and 9\%, respectively.

The FR3 source-follower PA in Fig.~\ref{fig:source_follower_pa} demonstrates an alternative low-\(|S_{22}|\) approach in 22-nm FDX+~\cite{eleraky_204_2026}. The PA uses a source follower as the final stage with \(C_{\mathrm{gs}}\) neutralization for gain and stability. The core area is 0.166~\(\mathrm{mm}^{2}\). Under a nominal 50-\(\Omega\) load, it reports a small-signal \(S_{22}<-10\)~dB from 11.5 to 25~GHz. At 12.25~GHz, it achieves \(P_{\mathrm{sat}}=16.9\)~dBm, \(\mathrm{OP}_{1\mathrm{dB}}=15.2\)~dBm, and 41.2\% PAE. Across 11.25--14.5~GHz, \(P_{\mathrm{sat}}\) remains > 14.7~dBm, \(\mathrm{OP}_{1\mathrm{dB}}\) exceeds 12.3~dBm, and peak PAE exceeds 27.5\%.

Under a 3:1 VSWR circle with load angles from 0\(^{\circ}\) to 315\(^{\circ}\) in 45\(^{\circ}\) steps~\cite{eleraky_204_2026}, the measured PG deviation is below 2~dB, the \(\mathrm{OP}_{1\mathrm{dB}}\) deviation remains below 2.97~dB, and the peak-PAE deviation is below 7.8\% relative to the matched case, across 12.5--14.5~GHz. Under 4:1 VSWR, the corresponding PG, \(\mathrm{OP}_{1\mathrm{dB}}\), and peak-PAE deviations remain below 2.25~dB, 3.5~dB, and 11.7\%, respectively.

With a 100~MSym/s 64-QAM signal and 8.5-dB PAPR under 3:1 and 4:1 VSWR circles, the PA maintains \(\mathrm{EVM}_{\mathrm{rms}}\) between approximately \(-27\) and \(-24.5\)~dB, while achieving \(P_{\mathrm{avg}}>10\), 10.5, and 8.5~dBm and \(\mathrm{PAE}_{\mathrm{avg}}>18.5\%\), 14.7\%, and 11\% at 12.5, 13.5, and 14.5~GHz, respectively. With a 12.5-GHz main tone and a 12.6-GHz reverse-coupled tone injected 4--6~dB below the PA output, the upper and lower \(\mathrm{RIMD}_{3}\) products are below approximately \(-48.25\) and \(-27.5\)~dBc, respectively.

Fig.~\ref{fig:loadpull_output_match_comparison}(a) compares the large-signal load-pull contours of the conventional standard-neutralized cascode PA with those of the complex-cascode LC-neutralized and source-follower PAs. The conventional cascode exhibits relatively narrow \(P_{\mathrm{sat}}\)/PAE contours, indicating stronger sensitivity to load-impedance variation. In comparison, the complex-LC PA provides a broader high-\(P_{\mathrm{out}}\)/high-PAE load region while bringing the output-matching condition closer to the load-pull optimum. The source follower provides a complementary low-output-impedance approach with broad \(\mathrm{OP}_{1\mathrm{dB}}\) and \(\mathrm{PAE}_{1\mathrm{dB}}\) contours.

Fig.~\ref{fig:loadpull_output_match_comparison}(b) further shows that the simulated LSSP \(S_{22}\) trajectories of both output-matched configurations remain within compact regions as \(P_{\mathrm{in}}\) increases. The complex-LC trajectory remains near the center of the Smith chart, while the source-follower trajectory remains compact over the reported 10--25-GHz frequency range. Both remain approximately within the 2:1 VSWR contour over the reported drive levels. These results indicate that low output reflection is largely preserved beyond the small-signal regime while maintaining useful large-signal load-pull performance~\cite{eleraky_55_2025, eleraky_compact_2026, eleraky_204_2026}.

\begin{figure}[t!]
    \centering
    \includegraphics[width=0.8\linewidth]{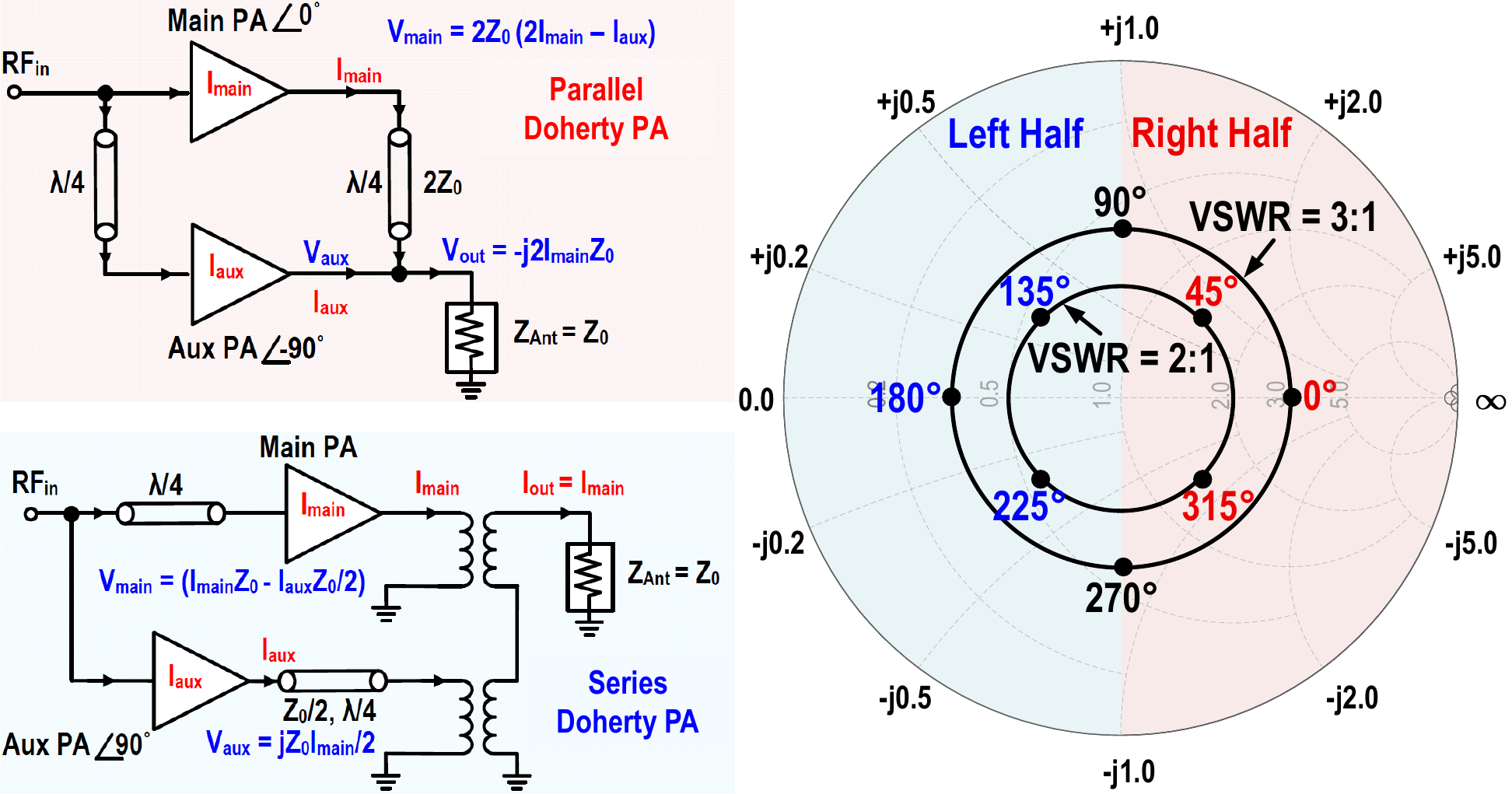}
    \caption{Parallel and series Doherty PA configurations and their complementary Smith-chart regions under antenna VSWR~\cite{hu_antenna_2015,mannem_reconfigurable_2020}.}
    \label{fig:Parallel_Series_Doherty_VSWR}
    \vspace{-0.5em}
\end{figure}

\begin{figure}[t!]
    \centering
    \includegraphics[width=0.9\linewidth]{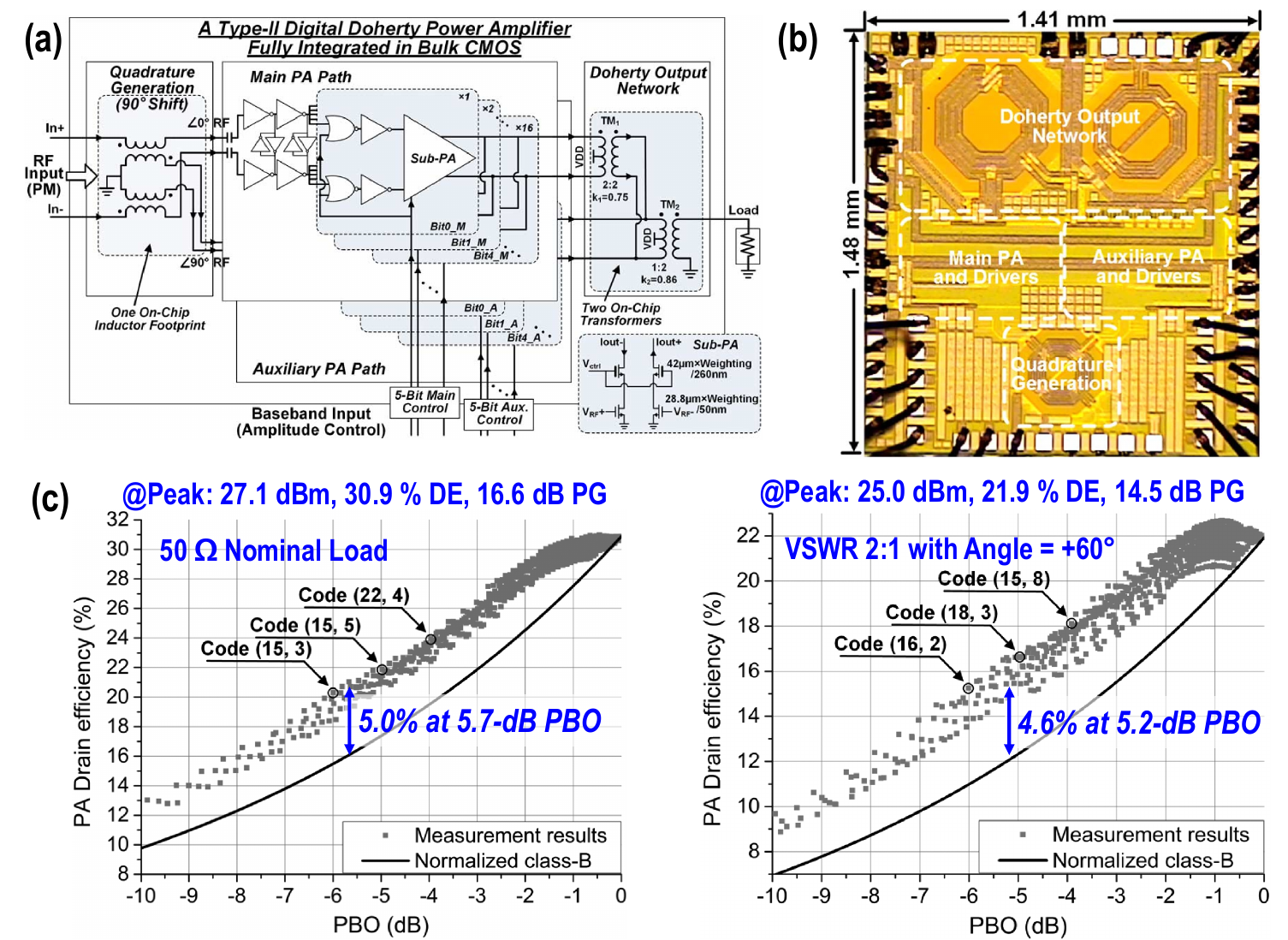}
    \caption{A 3.6-GHz Type-II digital Doherty PA in 65-nm bulk CMOS~\cite{hu_antenna_2015}. (a) Circuit diagram. (b) Chip micrograph. (c) Measured drain efficiency versus PBO at the nominal \(50~\Omega\) load and a 2:1 VSWR load with \(\angle\Gamma_{\mathrm{ant}}=+60^{\circ}\).}
    \label{fig:TypeII_Digital_DohertyPA}
    \vspace{-0.5em}
\end{figure}

\begin{figure}[t!]
    \centering
    \includegraphics[width=0.85\linewidth]{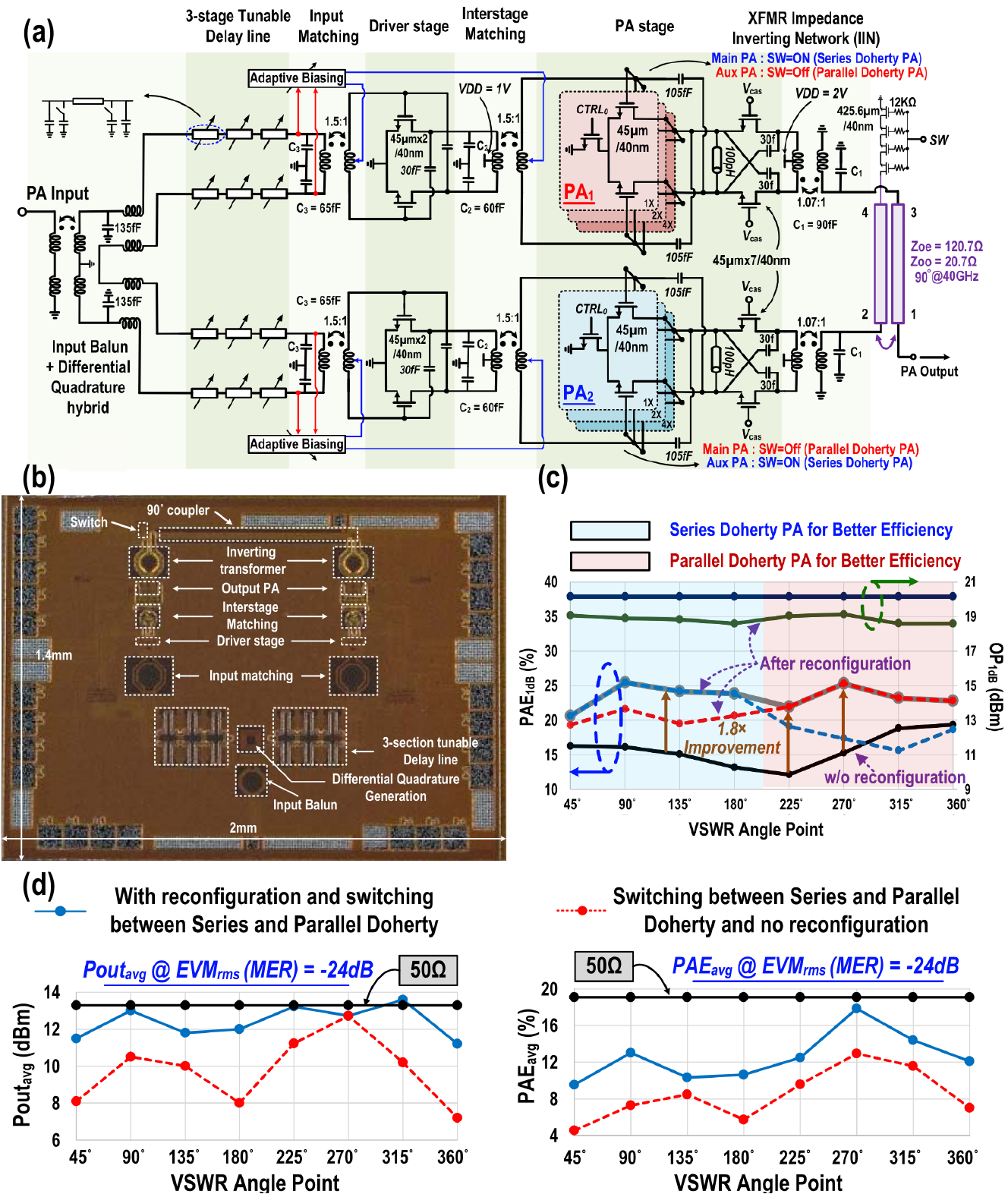}
    \caption{A 39-GHz reconfigurable hybrid series/parallel Doherty PA in 45-nm RFSOI CMOS~\cite{mannem_reconfigurable_2020}. (a) Circuit diagram. (b) Chip micrograph. Measured performance over a full-angle 3:1 VSWR circle: (c) \(\mathrm{PAE}_{1\mathrm{dB}}\) and \(\mathrm{OP}_{1\mathrm{dB}}\); and (d) \(P_{\mathrm{avg}}\) and \(\mathrm{PAE}_{\mathrm{avg}}\).}
    \label{fig:Role_Exchange_VSWR_DohertyPA}
    \vspace{-1.0em}
\end{figure}

\subsection{Back-Off Efficiency Enhanced PAs Under VSWR}

The preceding demonstrations mainly employ Class-AB/B PAs. With high-PAPR QAM and OFDM signals, conventional linear PAs operate predominantly at output power back-off (PBO), where their PAE decreases rapidly, reducing the average TX efficiency and increasing the thermal density of large arrays~\cite{wang_millimeter-wave_2021,pashaeifar_thesis_2024}. Doherty and other load-modulated PAs enhance PBO efficiency by modulating the effective load so that the Main PA continues to operate near its efficient voltage or current swing. Under antenna VSWR, the output combiner no longer presents the intended load-modulation trajectories, perturbing the branch loading, relative phase, and compression behavior. PBO efficiency enhancement and VSWR resilience should be co-designed~\cite{hu_antenna_2015,mannem_reconfigurable_2020,pashaeifar_millimeter-wave_2021,diverrez_22-44_2024,diverrez_theoretical_2026}. Fig.~\ref{fig:Parallel_Series_Doherty_VSWR} illustrates the Main/Aux PA configurations of parallel and series Doherty PAs. Owing to their dual load-modulation relationships, the parallel mode is preferred for the right-side Smith-chart region, \(\lvert Z_{\mathrm{ant}}\rvert>Z_0\), whereas the series mode is preferred for the left-side region, \(\lvert Z_{\mathrm{ant}}\rvert<Z_0\). This complementary behavior enables recovery of \(\mathrm{OP}_{1\mathrm{dB}}\) with the more favorable peak efficiency across both regions~\cite{hu_antenna_2015,mannem_reconfigurable_2020}.

Three digital Doherty configurations with different degrees of freedom in the carrier/peaking RF currents and relative phase are analyzed in~\cite{hu_antenna_2015}. These degrees of freedom allow the PA state to be selected for a target \(P_{\mathrm{out}}\) while avoiding carrier-amplifier voltage clipping and maximizing drain efficiency under the specified load condition. The fabricated Type-II digital Doherty PA employs two 5-bit RF power DACs, providing 1024 carrier/peaking RF-current configurations without additional phase tuning. Fabricated in 65-nm bulk CMOS, the prototype occupies \(1.48\times1.41~\mathrm{mm}^{2}\). At 3.6~GHz and a nominal \(50~\Omega\) load, it delivers a peak \(P_{\mathrm{out}}\) of 27.1~dBm with 30.9\% DE and provides an absolute efficiency improvement of 5.0\% over normalized Class-B operation at 5.7-dB PBO. At a selected 2:1 VSWR load with \(\angle\Gamma_{\mathrm{ant}}=+60^{\circ}\), it delivers a peak \(P_{\mathrm{out}}\) of 25.0~dBm with 21.9\% DE and provides an absolute efficiency improvement of 4.6\% at 5.2-dB PBO. Without DPD, 1-MSym/s QPSK and 500-kSym/s 16-QAM signals at \(50~\Omega\) achieve \(P_{\mathrm{avg}}=23.3/21.9\)~dBm, DEs of 22.9/18.2\%, \(\mathrm{EVM}_{\mathrm{rms}}=-29.1/-28.2\)~dB, and ACLR of \(-33.4/-35.3\)~dBc, respectively. Using load-dependent efficiency-optimum code sets across three selected 2:1 VSWR states with \(\lvert Z_{\mathrm{ant}}\rvert>Z_0\), both signals maintain \(\mathrm{EVM}_{\mathrm{rms}}<-25\)~dB and ACLR below \(-30\)~dBc. At the \(\angle\Gamma_{\mathrm{ant}}=+60^{\circ}\) state, the measured \(P_{\mathrm{avg}}\) values are 21.1/20.0~dBm with \(\mathrm{EVM}_{\mathrm{rms}}=-26.6/-25.8\)~dB and ACLR of \(-33.5/-36.0\)~dBc, corresponding to relative average efficiency improvements of 25\%/36\% over normalized Class-B operation.

\begin{figure}[t!]
    \centering
    \includegraphics[width=0.8\linewidth]{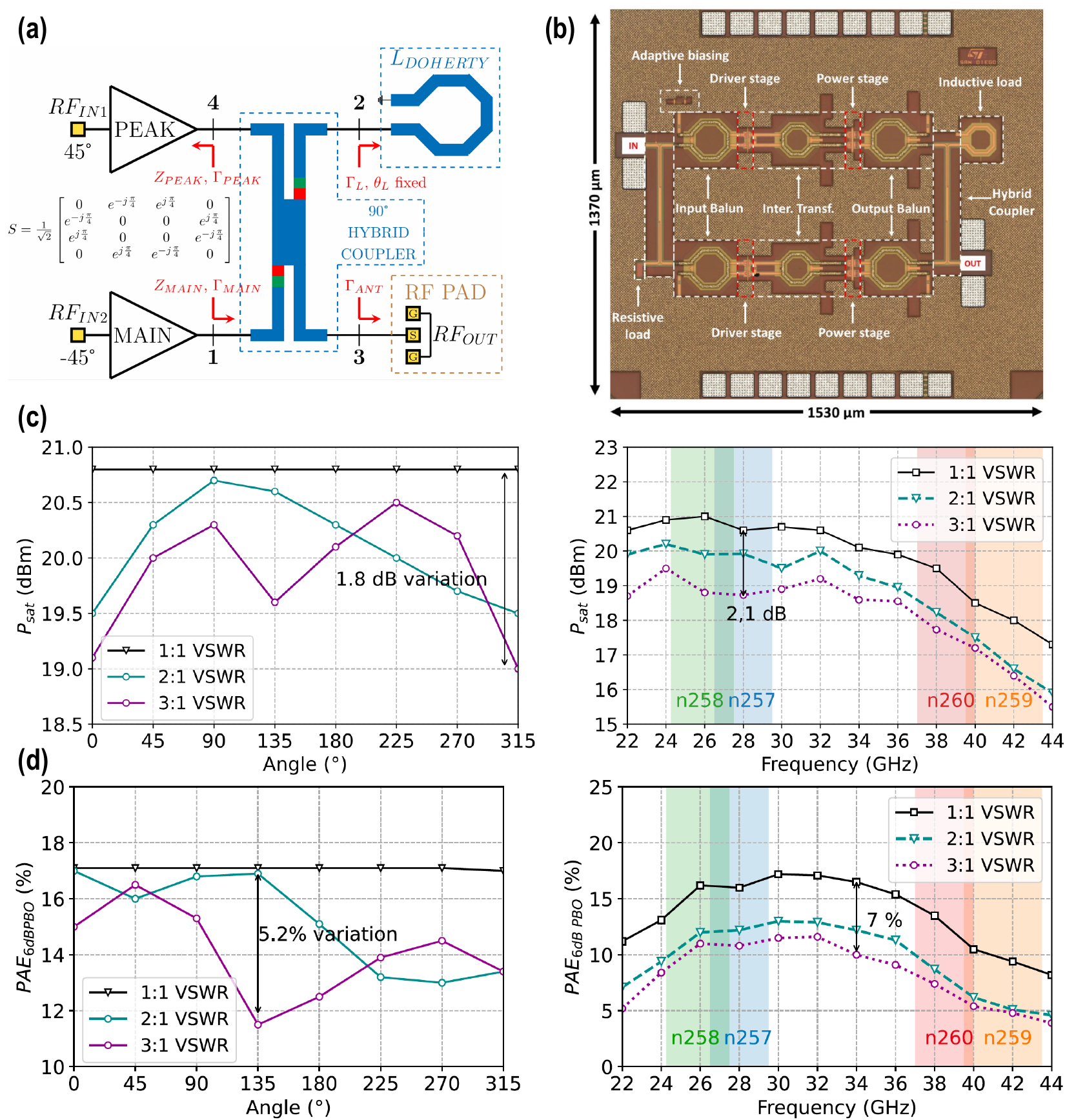}
    \caption{A 22--44-GHz quasi-balanced Doherty PA with an inductively terminated output hybrid in 28-nm FD-SOI CMOS~\cite{diverrez_22-44_2024,diverrez_theoretical_2026}. (a) Circuit diagram. (b) Chip micrograph. Measured results over a full-angle 3:1 VSWR circle: (c) \(P_{\mathrm{sat}}\) and (d) PAE at 6-dB PBO.}
    \label{fig:Quasi_Balanced_DohertyPA}
    \vspace{-1.0em}
\end{figure}

\begin{figure}[t!]
    \centering
    \includegraphics[width=0.9\linewidth]{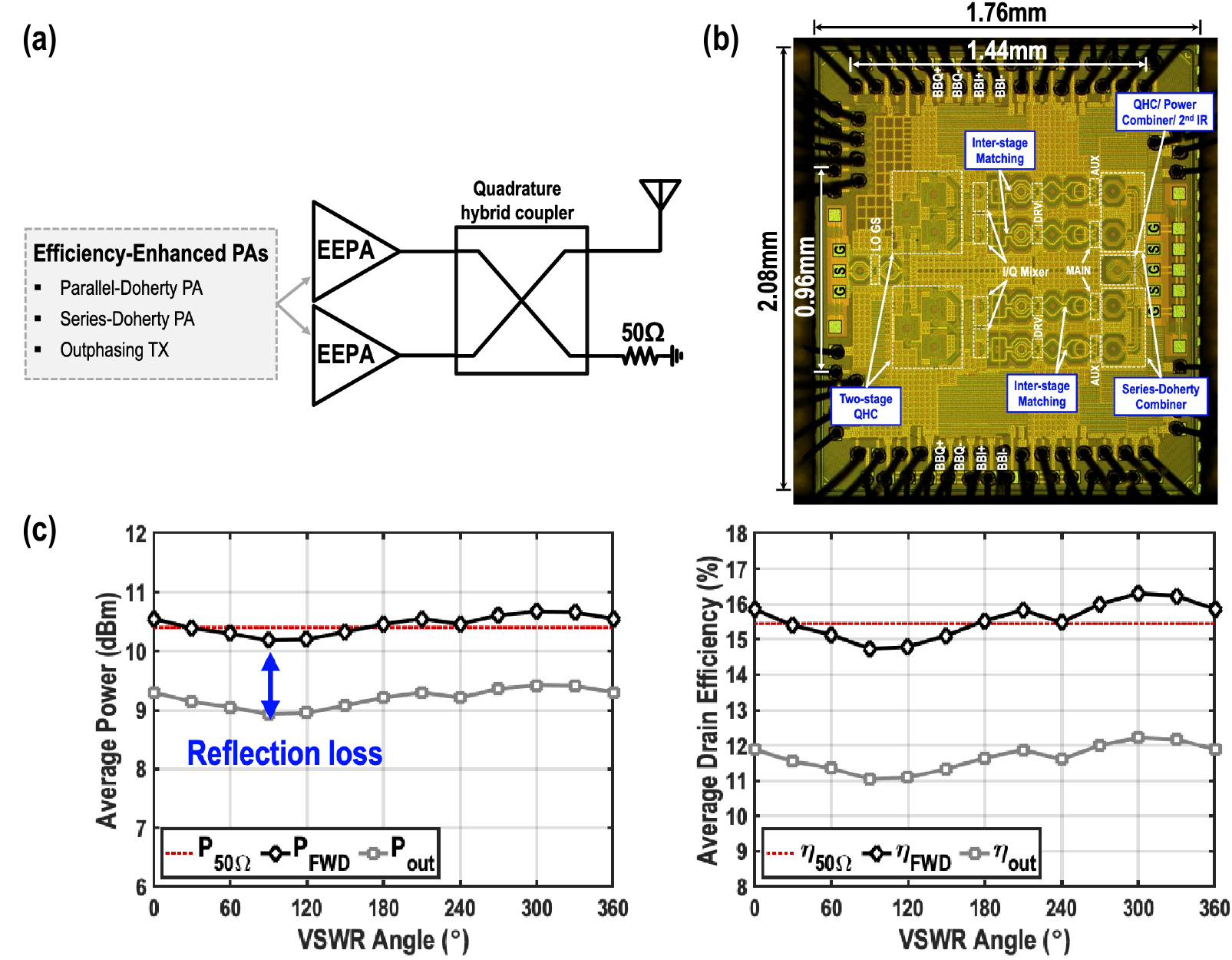}
    \caption{A 24--30-GHz mutual-coupling-resilient double-quadrature TX with an efficiency-enhanced balanced PA in 40-nm CMOS~\cite{pashaeifar_144_2021,pashaeifar_millimeter-wave_2021}. (a) Circuit diagram. (b) Chip micrograph. (c) Measured average \(P_{\mathrm{FWD}}\), \(P_{\mathrm{out}}\), and their corresponding drain efficiencies over a full-angle 3:1 VSWR circle.}
    \label{fig:Double_Quadrature_TX}
    \vspace{-0.5em}
\end{figure}

\begin{figure}[t!]
    \centering
    \includegraphics[width=0.9\linewidth]{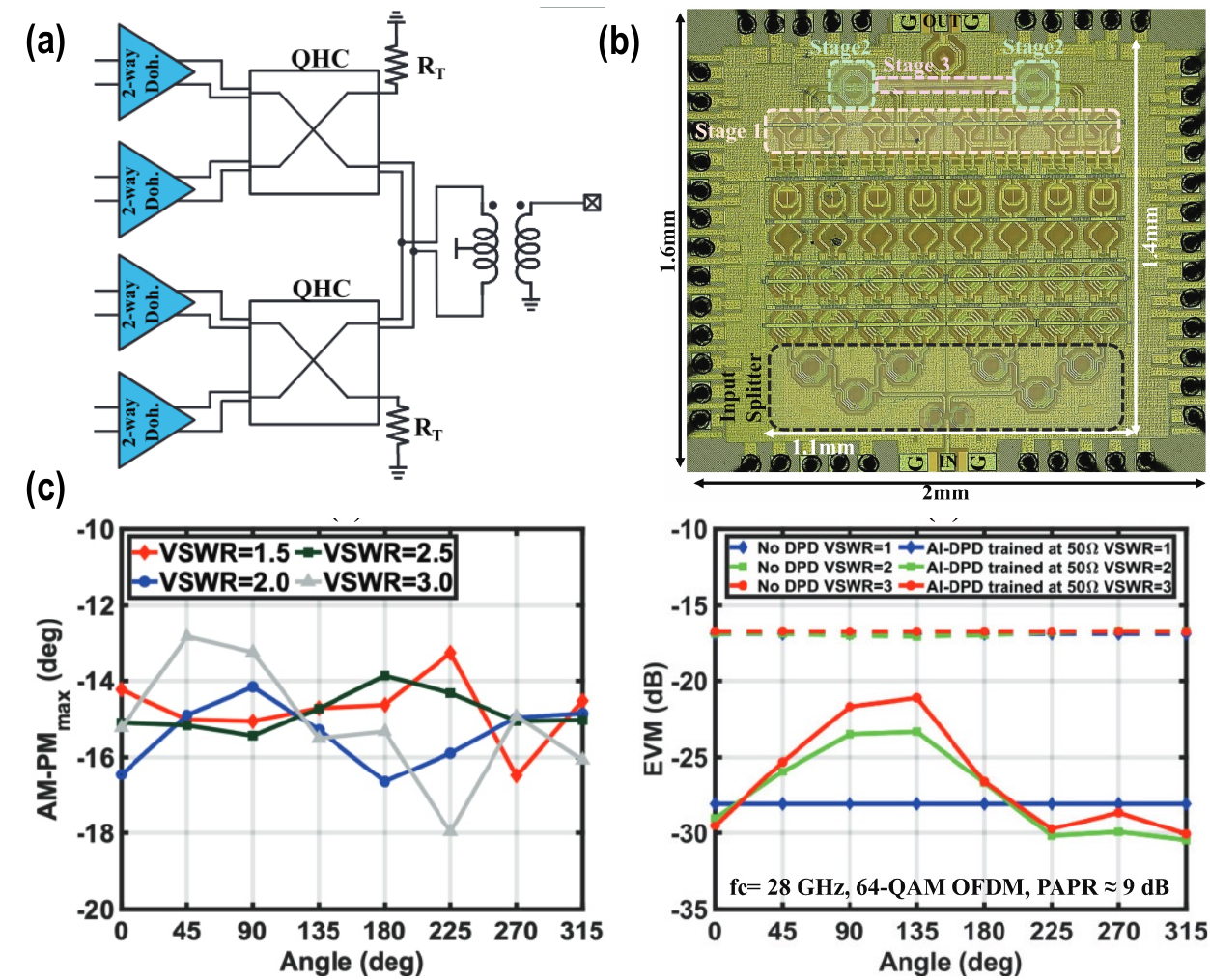}
    \caption{A 26--32-GHz \(4\times\) two-way Doherty PA in 40-nm CMOS~\cite{kumaran_4two-way_2025}. (a) Circuit diagram. (b) Chip micrograph. (c) Measured \(\mathrm{EVM}_{\mathrm{rms}}\) over a full-angle 3:1 VSWR circle.}
    \label{fig:FourByTwo_DPA}
    \vspace{-1.0em}
\end{figure}

A reconfigurable PA uses a four-port 90\(^{\circ}\)-coupler-based output network whose isolation port is switched between open and short terminations to realize parallel and series Doherty active load modulation, respectively, with the Main/Aux PA roles exchanged between the two modes~\cite{mannem_reconfigurable_2020} (Fig.~\ref{fig:Role_Exchange_VSWR_DohertyPA}(a)). For each antenna load, the Main-PA current strength sets the recovered \(P_{\mathrm{sat}}\), while the Aux-PA current strength and relative phase are configured to restore the Main-PA load toward its optimum and suppress early compression. The prototype occupies a \(2.86~\mathrm{mm}^{2}\) die area and a 1.18-\(\mathrm{mm}^{2}\) core area. At a nominal \(50~\Omega\) load and 39~GHz, it achieves \(P_{\mathrm{sat}}=20.8\)~dBm with 33.3\% peak PAE and \(\mathrm{OP}_{1\mathrm{dB}}=20.2\)~dBm with \(\mathrm{PAE}_{1\mathrm{dB}}=32.2\%\). After reconfiguration over a full-angle 3:1 VSWR circle, \(\mathrm{OP}_{1\mathrm{dB}}\) is recovered to 18.5--19.1~dBm with \(\mathrm{PAE}_{1\mathrm{dB}}\) of 20.6--25.3\%. With a 100~MSym/s single-carrier 64-QAM signal, the PA maintains \(P_{\mathrm{avg}}>11.2\)~dBm and \(\mathrm{PAE}_{\mathrm{avg}}>9.6\%\) at \(\mathrm{EVM}_{\mathrm{rms}}=-24\)~dB over the same circle. The reported prototype focuses on RF reconfiguration and applies preselected load-dependent states. Integrating an antenna-load sensor and online controller would extend the demonstrated selected-state recovery to autonomous closed-loop operation.

A 26--42-GHz multi-port active-load-pulling PA in 65-nm CMOS uses code-controlled interactions between PA branches to synthesize favorable impedances over frequency, PBO, and antenna load~\cite{chappidi_multi-port_2020}. The prototype delivers > 19-dBm \(P_{\mathrm{sat}}\) and > 20\% peak PAE over 28--40~GHz, reaching 24\% peak PAE at 33~GHz. Under selected load states up to 4:1 VSWR at 33~GHz, the measured peak-power degradation is approximately 2~dB while Doherty-like PBO efficiency enhancement is retained. Because the accessible loads were limited by the external interface, this result represents selected-load validation instead of a full-angle VSWR-circle sweep.

The quasi-balanced Doherty PA performs load modulation within a coupler-based balanced structure and uses an inductive termination to preserve the desired main-path impedance transformation~\cite{diverrez_22-44_2024,diverrez_theoretical_2026} (Fig.~\ref{fig:Quasi_Balanced_DohertyPA}(a)). The 28-nm FD-SOI CMOS prototype occupies a 0.82-\(\mathrm{mm}^{2}\) core area. At 28~GHz, it achieves \(P_{\mathrm{sat}}=20.3\)~dBm, 34.4\% peak PAE, 24\% PAE at 6-dB PBO, and 16\% PAE at 9.7-dB PBO. Across 22--44~GHz, \(P_{\mathrm{sat}}\), peak PAE, and PAE at 6-dB PBO remain > 17.3~dBm, 17.5\%, and 12\%, respectively. The full-angle 3:1 VSWR circle is sampled at 45\(^{\circ}\) increments. Relative to the matched case, the verified worst-case reductions over 22--44~GHz are less than 2.1~dB in \(P_{\mathrm{sat}}\), less than 11 percentage points in peak PAE, and less than 7 percentage points in PAE at 6-dB PBO. The reported VSWR validation is CW, while modulated-signal operation under VSWR is not reported.

The double-quadrature TX combines a direct upconverter with an efficiency-enhanced balanced PA (Fig.~\ref{fig:Double_Quadrature_TX}(a)), in which two series-Doherty branches are combined through a quadrature hybrid~\cite{pashaeifar_144_2021,pashaeifar_millimeter-wave_2021}. The measured TX output reflection coefficient is below \(-18\)~dB from 22.5 to 30~GHz and reaches \(-22.2\)~dB at 27~GHz. The prototype occupies a 1.38-\(\mathrm{mm}^{2}\) core area and delivers \(\mathrm{OP}_{1\mathrm{dB}}\) of approximately 20~dBm, with 40\% and 31\% drain efficiency at \(\mathrm{OP}_{1\mathrm{dB}}\) and 6-dB PBO, respectively. Without digital predistortion, an eight-carrier 100-MHz 64-QAM OFDM signal with 800-MHz aggregated bandwidth achieves an average \(P_{\mathrm{out}}\) of 8.4~dBm, 10.8\% drain efficiency, and EVM better than \(-27.1\)~dB. With a 100-MHz 64-QAM signal over a full-angle 3:1 VSWR circle, the maximum 27-GHz deviations are 0.3~dB in \(P_{\mathrm{FWD}}\)/gain, 1.65~dB in EVM, and 1~dB in ACLR. At 28~GHz, the maximum deviations are 0.65~dB in \(P_{\mathrm{FWD}}\)/gain and 3.9~dB in EVM. Following the power accounting in Section~II-B, \(P_{\mathrm{out}}\) includes the passive accepted-power factor \(1-|\Gamma_{\mathrm{ant}}|^2\), whereas the small \(P_{\mathrm{FWD}}\) variation shows suppression of the additional PA/TX reinteraction.

A 26--32-GHz \(4\times\) two-way Doherty PA in 40-nm CMOS combines four two-way Doherty cells using two quadrature hybrid couplers and a balun~\cite{kumaran_4two-way_2025}. At 28~GHz, it achieves 25.2-dBm \(P_{\mathrm{sat}}\), 25.5-dB power gain, and 20.5\%/13.3\% drain efficiency at peak/6-dB PBO, while supporting multi-Gb/s OFDM signals. Across the measured 2:1 and 3:1 VSWR load states, the quadrature-coupler-based output network keeps gain and \(P_{1\mathrm{dB}}\) deviations below approximately 1 and 0.8~dB, respectively. A DPD trained at the nominal \(50~\Omega\) load further improves the modulated-signal EVM under mismatch. This work combines power combining, PBO efficiency enhancement, and measured VSWR resilience in a high-output-power silicon implementation.

\subsection{Representative III--V Implementations}
\begin{figure}[!t]
    \centering
    \includegraphics[width=0.75\linewidth]{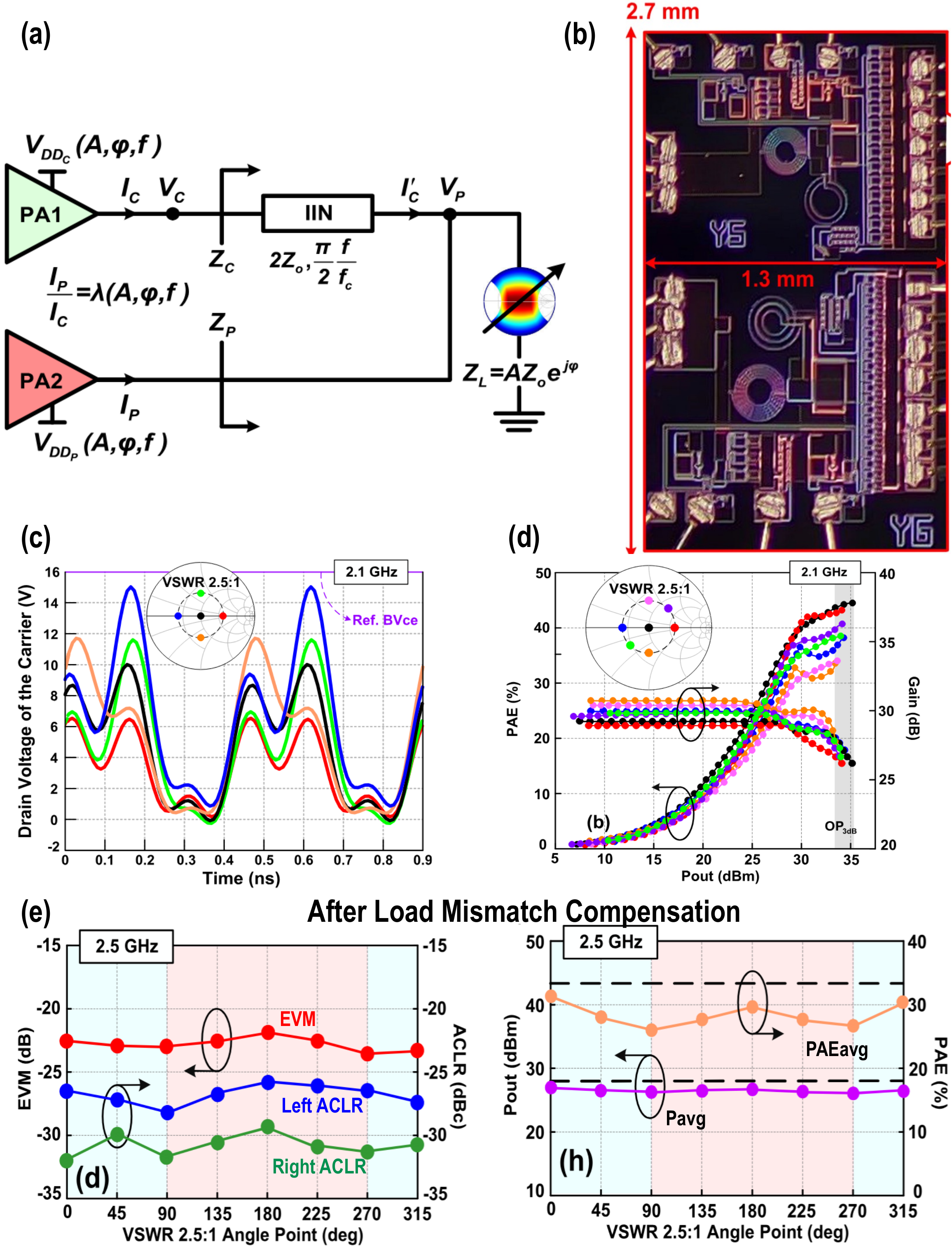}
    \caption{A 1.7--2.7-GHz GaAs-HBT dual-input Doherty PA MMIC~\cite{guo_ultra-broadband_2025}. (a) Circuit diagram. (b) Chip micrograph. (c) Simulated saturated carrier drain-voltage waveforms and (d) measured PAE and gain at 2.1~GHz over a 2.5:1 VSWR circle. (e) Measured EVM, ACLR, \(P_{\mathrm{avg}}\), and \(\mathrm{PAE}_{\mathrm{avg}}\) over a 2.5:1 VSWR circle at 2.5~GHz after mismatch compensation.}
    \label{fig:Guo2025_GaAsDPA}
    \vspace{-0.5em}
\end{figure}

\begin{figure}[!t]
    \centering
    \includegraphics[width=0.75\linewidth]{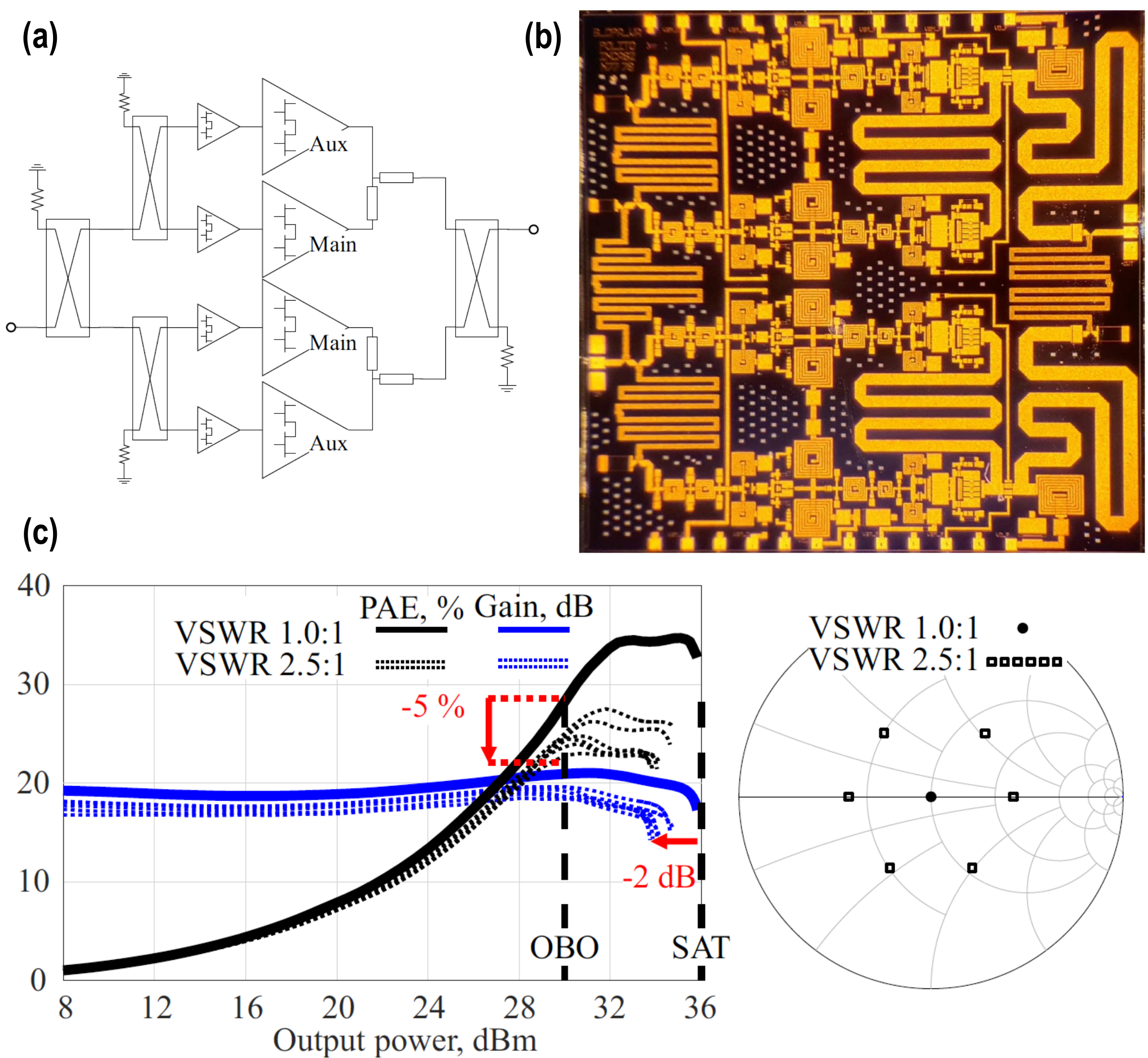}
    \caption{A 3.3--5.0-GHz, 3-W stacked GaAs-pHEMT balanced Doherty PA MMIC~\cite{piacibello_wideband_2025}. (a) Circuit diagram. (b) Chip micrograph. (c) Measured gain and PAE for the nominal load and selected 2.5:1 VSWR terminations.}
    \label{fig:Piacibello2025_GaAsDPA}
    \vspace{-1.0em}
\end{figure}

\begin{figure}[t!]
    \centering
    \includegraphics[width=0.94\linewidth]{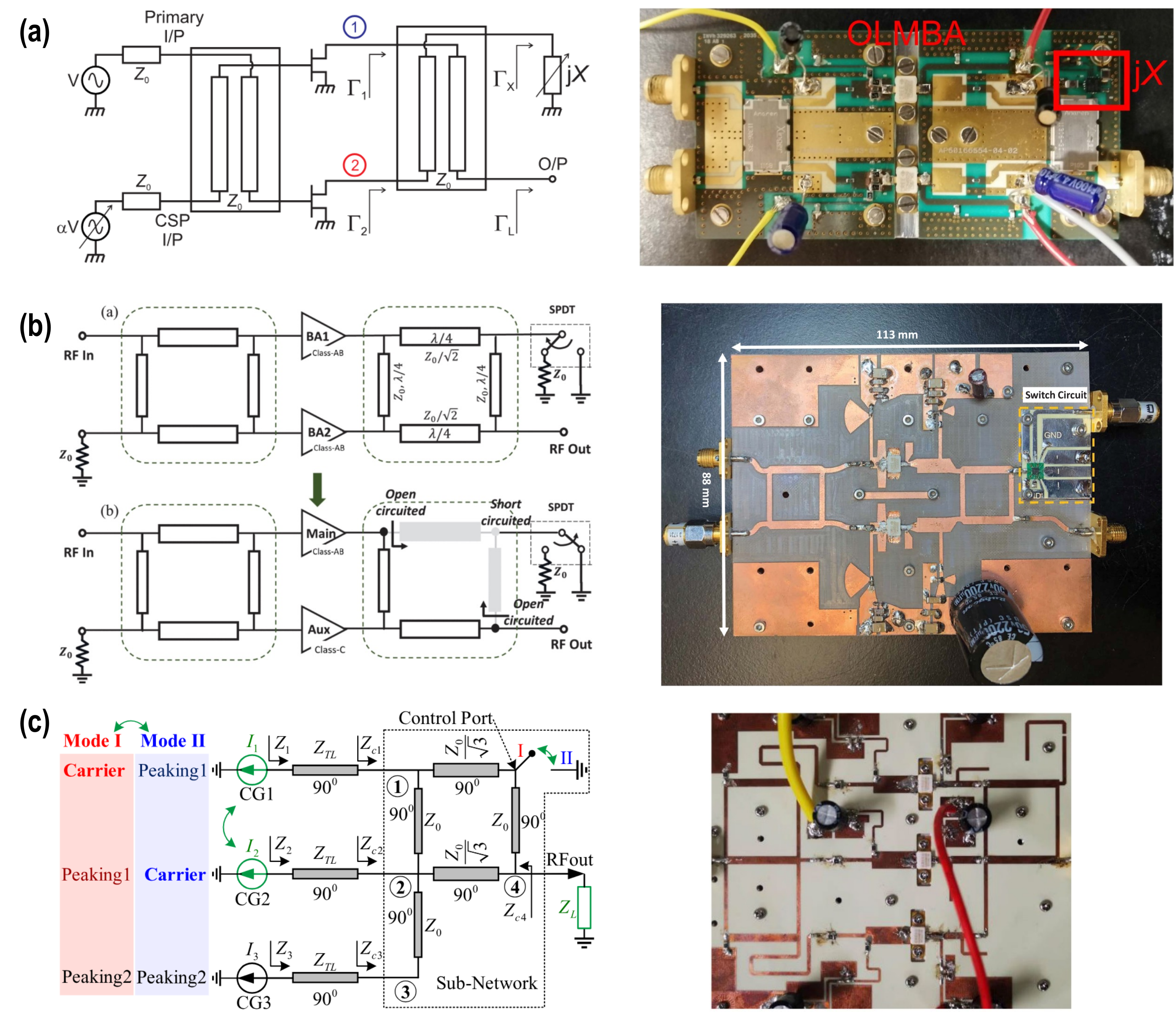}
    \caption{VSWR-resilient GaN-HEMT PAs at Sub-7 GHz. (a) Orthogonal load-modulated balanced amplifier (OLMBA)~\cite{quaglia_mitigation_2022}. (b) Balanced-to-Doherty (B2D) mode-reconfigurable PA~\cite{lyu_balanced--doherty_2020}. (c) Dual-mode three-way Doherty PA~\cite{pang_dual-mode_2024}.}
    \label{fig:board_level_schematics}
    \vspace{-1.0em}
\end{figure}

The silicon demonstrations reviewed above emphasize integration density and array-element footprint.
III--V MMICs provide an alternative in which higher breakdown voltage and power gain support larger output swing and higher optimum load impedance. The relevant comparison criteria are the VSWR mitigation mechanism and the verified load region.

A GaAs-HBT Doherty PA (DPA) combines feedback and feedforward dual-adaptive biasing at 2.5~GHz, enabling a 100-MHz 5G signal without DPD at 42.8\% PAE and \(-35~\mathrm{dBc}\) ACLR under nominal loading while maintaining ACLR below \(-30~\mathrm{dBc}\) on a 2:1 VSWR circle~\cite{imai_dual-adaptive_2025}.

A 1.7--2.7-GHz dual-input DPA in Fig.~\ref{fig:Guo2025_GaAsDPA} is fabricated in a 250-nm GaAs-HBT process and adapts the carrier and peaking supply voltages and relative input amplitude and phase for load-mismatch compensation~\cite{guo_ultra-broadband_2025}. The simulated saturated-carrier drain voltage remains below \(BV_{\mathrm{CE}}\) over a 2.5:1 VSWR circle, while compensated CW measurements at 2.1~GHz limit the output-power loss to 1.4~dB and the PAE degradation to 8\% at 6-dB PBO and 11\% at saturation. At nominal load, the peak PAE is 39--44.5\%, with \(P_{\max}\) of 34.2--35.1~dBm. Over the compensated 2.5:1 VSWR circle, the peak PAE is 31.5--43\%, with \(P_{\max}\) of 33--34.4~dBm, and the 6-dB-PBO PAE is 30--38\%. With a 100-MHz 64-QAM signal having 6.5-dB PAPR, \(P_{\mathrm{out,avg}}\) is 26.5--27.9~dBm and \(\mathrm{PAE}_{\mathrm{avg}}\) is 26.5--35\%, with EVM below \(-22\)~dB and ACLR below \(-25.5\)~dBc.

A 3.3--5.0-GHz 3-W balanced DPA in Fig.~\ref{fig:Piacibello2025_GaAsDPA} combines two Doherty cells and employs stacked devices in the driver and final stages together with integrated quadrature Lange couplers in a 150-nm GaAs-pHEMT process~\cite{piacibello_wideband_2025}. The measured \(S_{11}\) and \(S_{22}\) remain below \(-15\)~dB over the operating band, while \(P_{\mathrm{sat}}\) exceeds 34.7~dBm and the PAE at 6-dB PBO exceeds 25\%. Under selected 2.5:1 VSWR terminations at 4.7~GHz, the worst-case \(P_{\mathrm{sat}}\) reduction is $\sim$ 2~dB, and the 6-dB-PBO PAE decreases from 28\% to 23\%.

At sub-7-GHz frequencies, GaN HEMTs can provide the voltage swing and power capability required for multi-watt-level VSWR-resilient PAs. These PAs mainly employ balanced combining, active load modulation, mode reconfiguration, or load-dependent supply and input control to maintain output power and efficiency under antenna VSWR variation, as shown in Fig.~\ref{fig:board_level_schematics}. In the orthogonal load-modulated balanced amplifier (OLMBA), the complex control-signal ratio and a reactive termination at the isolated port set the effective branch loads~\cite{quaglia_mitigation_2022}. The balanced-to-Doherty architecture selects balanced or Doherty operation through the isolated-port termination and the carrier/peaking assignments~\cite{lyu_balanced--doherty_2020}. The dual-mode three-way Doherty PA similarly changes the carrier/peaking assignments and the control-port termination to preserve a wider high-efficiency power range under antenna VSWR variation~\cite{pang_dual-mode_2024}. These prototypes show the physical implementation of the couplers, switching networks, reactive terminations, bias interfaces, and high-power device arrangements required by the corresponding architectures.

\section{Future Challenges and Opportunities}

Mm-Wave and cm-Wave PAs are typically employed in phased-array and hybrid-beamforming systems. Each PA must provide sufficient $P_{\mathrm{out}}$ and efficiency within a $\lambda/2 \times \lambda/2$ antenna-element lattice, while the array-level thermal density, calibration complexity, and implementation overhead remain manageable as the number of elements increases. These coupled requirements correspond to the array energy, spatial-Nyquist, thermal, and complexity walls~\cite{wang_millimeter-wave_2021,eleraky_204_2026,eleraky_55_2025,eleraky_compact_2026, pashaeifar_thesis_2024,liu_advanced_2025}. This section identifies future research opportunities toward compact, VSWR-resilient, energy-efficient, high-power-density, and calibration-scalable integrated PAs.

\subsection{Ultra-Compact VSWR-Resilient PAs With Back-Off Efficiency Enhancement}

Spectrally efficient modulation schemes, including high-order QAM, OFDM, and carrier aggregation (CA), exhibit high peak-to-average power ratios (PAPRs). The PA must accommodate the peak signal power while its average output power remains several decibels below $P_{\mathrm{sat}}$. Even an ideal Class-B PA with a lossless output network decreases from 78.5\% drain efficiency at peak power to $\sim$ 39.3\% at 6-dB PBO, while practical Class-A/AB PAs generally exhibit a larger reduction~\cite{huang_261_2021,huang_35100ghz_2021,huang_coupler_2023,mannem_reconfigurable_2020,mannem_broadband_2021,liu_3210_2024,liu_329_2024,liu_broadband_2024,liu_advanced_2025}. High PBO and average efficiency are essential for energy-efficient array operation~\cite{wang_millimeter-wave_2021, mannem_reconfigurable_2020,mannem_broadband_2021, diverrez_22-44_2024,chappidi_multi-port_2020}, and the thermal behavior cannot be simply inferred from the efficiency of an isolated PA alone. The temperature of each element is influenced by the dissipation of neighboring channels and by the heat flow through the package and antenna aperture. Antenna VSWR further changes the PA dissipation and device stress as the beam state varies. Future array analysis should include VSWR-dependent PA efficiency and reliability within a joint electronics--package--antenna electromagnetic--thermal co-design~\cite{wang_millimeter-wave_2021,pashaeifar_thesis_2024}.

\begin{figure}[t!]
    \centering
    \includegraphics[width=0.8\linewidth]{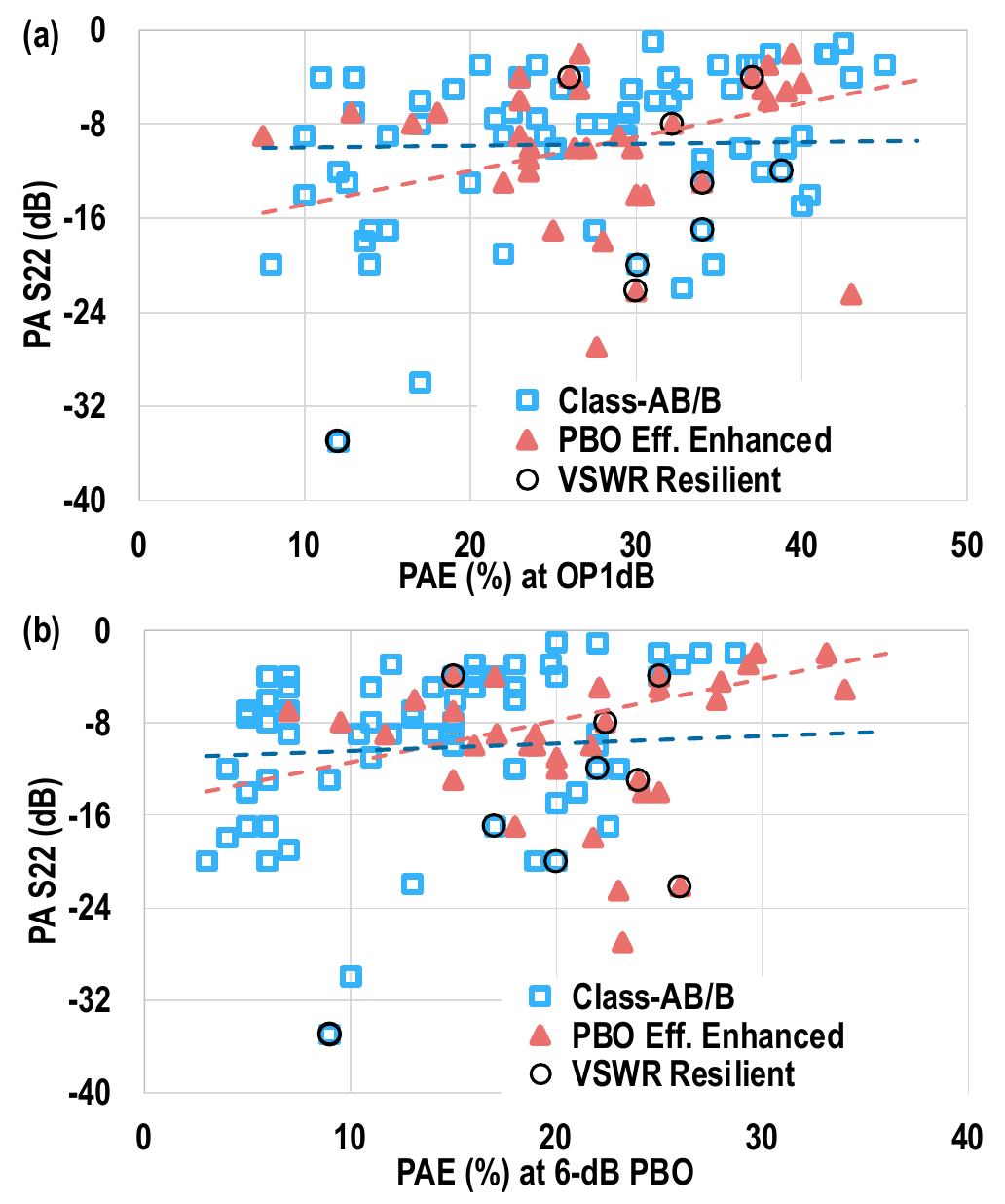}
    \caption{Small-signal \(S_{22}\) versus (a) \(\mathrm{PAE}_{1\mathrm{dB}}\) and (b) PAE at 6-dB PBO for state-of-the-art mm-Wave and cm-Wave PAs. Data are compiled from
   ~\cite{noauthor_pa_nodate}.}
    \label{fig:s22_pae_benchmark}
    \vspace{-1.0em}
\end{figure}

Figure~\ref{fig:s22_pae_benchmark}(a) examines whether improved output matching is compatible with competitive large-signal efficiency. Using $S_{22}<-8$~dB as a commonly adopted practical criterion in industry, the substantial overlap between the \(S_{22}<-8~\mathrm{dB}\) and \(S_{22}>-8~\mathrm{dB}\) groups shows that output-matched PAs do not necessarily result in poor $\mathrm{PAE}_{1\mathrm{dB}}$. Small-signal $S_{22}$ alone does not establish large-signal VSWR resilience, which should also be verified using load-pull or direct VSWR measurements. Nevertheless, the benchmark supports the feasibility of SOLM designs that combine output matching with the desired large-signal loadline~\cite{pecile_study_2025,eleraky_55_2025, eleraky_compact_2026}. Their fixed-state operation is attractive for large arrays because it can reduce the calibration effort associated with load-dependent state selection. Low $S_{22}$ can also reduce reverse excitation and the associated RIMD~\cite{atanasov_reverse_2020,eleraky_204_2026}.

At 6-dB PBO, Doherty and other load-modulated PAs occupy much of the higher-efficiency region in Fig.~\ref{fig:s22_pae_benchmark}(b). This motivates the joint design of VSWR resilience and active load modulation. The PA examples reviewed in Section~IV-E show that PBO efficiency enhancement and VSWR resilience can be achieved within the same PA architecture. Doherty PAs use multiple power paths and active load modulation, which often require additional combining networks and larger implementation area than compact Class-AB PAs~\cite{kumaran_26ghz_2023,kumaran_single-supply_2024,liu_broadband_2024,zhang_264_2021,zhang_millimeter-wave_2024,torii_efficiency_2024,urvoy_orthogonal_2026}. Meanwhile, translating LMBA and quasi-balanced Doherty architectures to integrated arrays can also require additional PA paths, couplers, and terminations, increasing area and passive loss~\cite{mannem_reconfigurable_2020,pashaeifar_144_2021,pashaeifar_millimeter-wave_2021,diverrez_22-44_2024,diverrez_theoretical_2026,guo_1-d_2023,urvoy_orthogonal_2026,chu_phase_2022,chu_investigation_2022,chu_broadband_2023,nikandish_unbalanced_2021}.

A major challenge is to realize an ultra-compact VSWR-resilient Doherty PA that satisfies the antenna-element area constraint. Recent works have compressed the Doherty output network into a single-transformer footprint. A 28-GHz parallel DPA uses an asymmetrically coupled transformer balun to integrate the impedance-inverting and impedance-scaling networks~\cite{liu_28-ghz_2026,liu_329_2024}. A 12-GHz series DPA similarly folds the series load-modulation network into a compact transformer-based combiner~\cite{xu_201_2026}. These demonstrations show that two-way Doherty PAs can approach the area of compact Class-AB implementations~\cite{park_single_2022,oh_322_2024, park_26--39ghz_2022,lee_54_2025, liu_28-ghz_2026,xu_201_2026,xu_compact_2025}.
A 25-GHz asymmetrical DPA further combines a single-transformer-footprint output network, a main path designed for simultaneous output and loadline matching (SOLM), and high-speed reconfigurable adaptive biasing~\cite{svelto_25-ghz_2026}. The PA supports data rates up to 12~Gb/s and reports an \mbox{$\sim5\times$} reduction in the chip core area relative to prior VSWR-resilient PBO efficiency-enhanced PAs.

Beyond two-way operation, three-way and higher-order $N$-way Doherty PAs can extend efficiency enhancement toward the 9--12-dB PBO region required by wideband OFDM signals~\cite{kumaran_single-supply_2024, liu_broadband_2024,zhang_millimeter-wave_2024, kumaran_4two-way_2025}. Extending compact output networks and VSWR-resilient operation to $N$-way architectures while preserving bandwidth, linearity, and low output reflection remains an important research opportunity.

\subsection{High-Power and High-Power-Density VSWR-Resilient PAs}

As the operating frequency increases, the available device power gain decreases because of finite $f_{\mathrm{T}}$ and $f_{\mathrm{max}}$, while the limited voltage handling of scaled silicon devices restricts the achievable PA output swing~\cite{wang_millimeter-wave_2021,camarchia_review_2020}. At a fixed antenna-element pitch, sufficient power gain, $P_{\mathrm{out}}$, and output-power density are required simultaneously to address the array energy wall within the available $\lambda/2\times\lambda/2$ element footprint.

\begin{figure}[t!]
    \centering
    \includegraphics[width=0.75\linewidth]{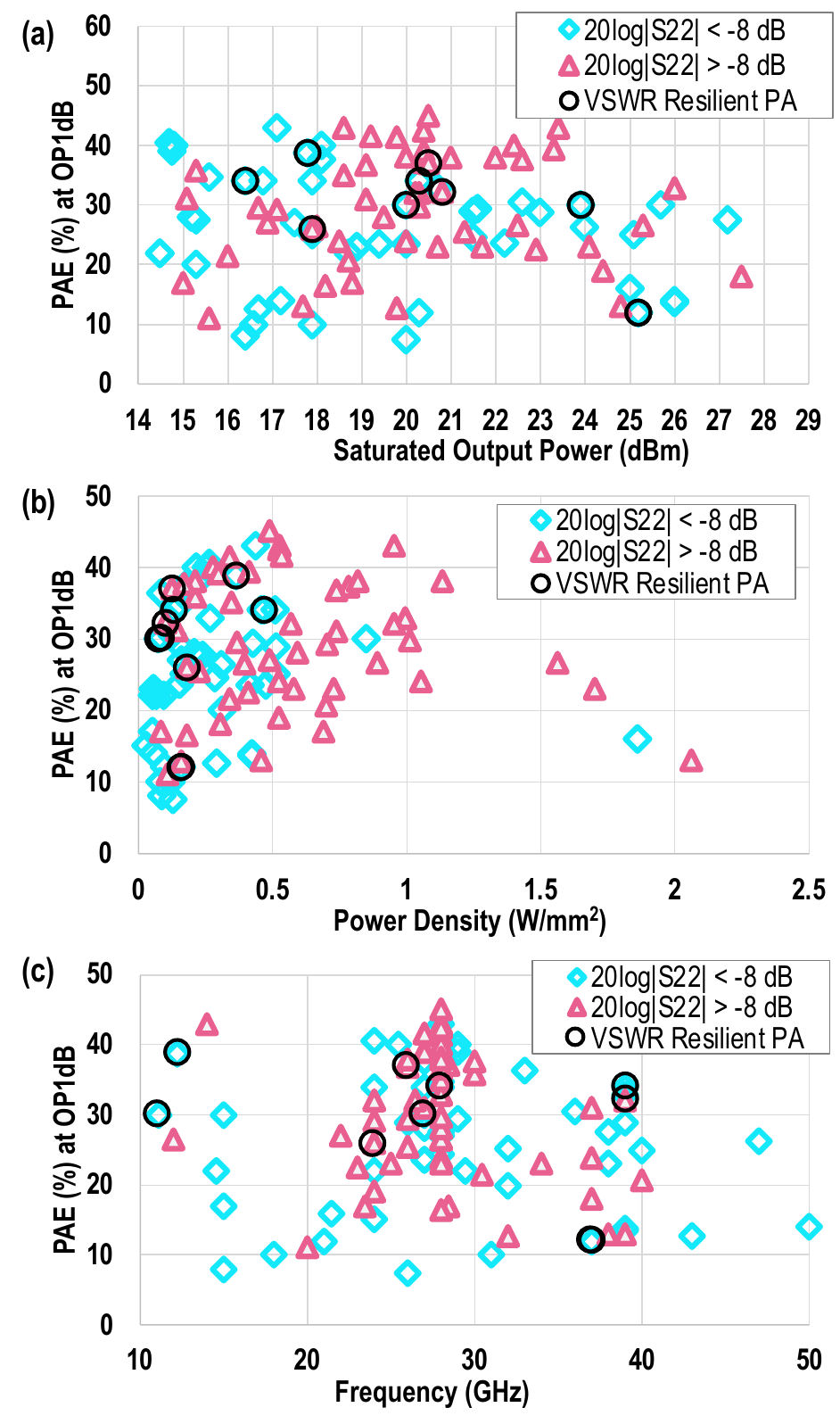}
    \caption{Benchmark of
     $\mathrm{PAE}_{1\mathrm{dB}}$ for 10--50~GHz silicon-integrated PAs versus
    (a) $P_{\mathrm{sat}}$,
    (b) output power density, and
    (c) carrier frequency.}
    \label{fig:freq_psat_benchmark}
    \vspace{-1.0em}
\end{figure}

Fig.~\ref{fig:freq_psat_benchmark}(a) shows that several VSWR-resilient silicon PAs reach the upper $P_{\mathrm{sat}}$ range of the present benchmark. Many of these high-$P_{\mathrm{sat}}$ demonstrations employ balanced or multi-way power combining, Doherty combiners, or isolator-assisted interfaces to combine power and maintain PA performance under antenna VSWR variation~\cite{pashaeifar_chain-weaver_2024, pashaeifar_millimeter-wave_2026, ghorbanpoor_332_2026}. These area and insertion-loss overheads prevent high $P_{\mathrm{sat}}$ from translating directly into high output-power density. This tradeoff is evident in Fig.~\ref{fig:freq_psat_benchmark}(b), where the highest output-power density among the VSWR-resilient silicon PAs is  $\sim 0.5~\mathrm{W/mm^2}$. The high-output-power-density region includes the output-matched source-follower and complex-LC cascode PAs, which embed VSWR resilience within compact final power stages~\cite{eleraky_55_2025,eleraky_compact_2026, eleraky_204_2026}.

Another central challenge is to scale output-matched PAs toward higher output power while preserving low $|S_{22}|$ and the desired large-signal loadline. Cascode and stacked PAs increase the available voltage swing and $P_{\mathrm{out}}$ in low-voltage silicon technologies~\cite{xu_compact_2025,xu_ultra-compact_2025, ghorbanpoor_12_2024,dabag_analysis_2013, kim_analysis_2015,oh_analysis_2026}. The resulting effective output impedance, parasitic capacitance, and inter-device voltage distribution increase the difficulty of maintaining low $|S_{22}|$, stability, and a low-loss output network. Future PA designs should co-optimize stacking, neutralization, device size, and the output matching network while preserving the desired large-signal loadline.

III--V technologies, including GaAs, GaN, and InP, provide a complementary path toward high-power VSWR-resilient PAs~\cite{camarchia_review_2020,pang_broadband_2022,torii_high-efficiency_2025,piacibello_wideband_2025,chu_3674-ghz_2026,huang_35100ghz_2021,chu_10-16_2025}. The high breakdown voltage supports a large output swing and a high optimum load impedance, relaxing the required impedance transformation. Besides the VSWR-resilient PA examples in Section IV-F, GaN MMIC PA research spans sub-7-GHz, cm-Wave, mm-Wave, and D-band operation, including distributed, Doherty, and power-combined architectures~\cite{lv_dual_band_gan_2019,lv_distributed_gan_2021,lv_highly_linear_doherty_2022, lv_distributed_adaptive_2024, piacibello_three_way_2023, piacibello_balanced_stacked_2023, giofre_two_way_gan_2024,piacibello_high_gain_2024, piacibello_npr_2025,ramella_space_grade_2025, fang_linear_wideband_2025,shi_wideband_mmwave_2023, shi_classf_gan_2023,li_dband_gan_2025}. These studies mainly emphasize output power, bandwidth, PBO efficiency, linearity, and integration. Future III--V PA demonstrations should report the $S_{22}$, device voltage and current stress, and modulated-signal performance under full-angle VSWR conditions.

Combining the advantages of III--V devices with silicon's high integration density and extensive analog, mixed-signal, and digital functionality has motivated the co-integration of multiple device technologies through advanced packaging and common-substrate approaches. GaN-on-silicon growth is already widely practiced, while recent demonstrations have employed CMOS-compatible processing of GaN-on-silicon circuits followed by wafer bonding to CMOS ICs~\cite{wang_millimeter-wave_2021}.

\subsection{Array-Scalable Operation, Calibration, and Intelligent Control}

Operating and calibrating VSWR-resilient PAs becomes increasingly complex as the number of array elements grows. The optimum PA state can vary with beam configuration and operating frequency and can drift with circuit and environmental conditions. Exhaustive per-element characterization and multidimensional lookup tables (LUTs) become increasingly impractical, slowing calibration and beam-state updates in large arrays~\cite{wang_millimeter-wave_2021, pashaeifar_thesis_2024,liu_advanced_2025}. Temperature-compensated biasing and stable gain and phase control can reduce this burden by allowing the same calibration settings to remain valid over a wider operating range.

Output-matched PAs can maintain performance in a fixed operating state, while reconfigurable PAs require the identification of a suitable state. Autonomous state identification remains a key array-level challenge~\cite{eleraky_55_2025,eleraky_204_2026, liu_24_2025,mannem_reconfigurable_2020}. An early 28-GHz CMOS self-healing PA established the feasibility of a fully integrated sensor-to-controller-to-actuator loop~\cite{bowers_integrated_2013}. Many recent VSWR-resilient PAs continue to rely on preselected states or offline control. Compact, fast, and calibration-scalable autonomous recovery under wideband modulated-signal operation remains insufficiently explored~\cite{munzer_single-ended_2021, munzer_single-ended_2022,munzer_broadband_2022, munzer_broadband_2023,liu_3210_2024, liu_2739-ghz_2025,zhang_264_2021}. Meanwhile, most VSWR-resilient PAs included in the present benchmark operate between 10 and 40~GHz (Fig.~\ref{fig:freq_psat_benchmark}(c)), leaving the upper mm-Wave and sub-THz frequency ranges largely unexplored.

Beyond PA-state selection, array-level calibration must also address load-dependent nonlinear distortion. Behavioral models derived from PA measurements or mm-Wave load-pull data and combined with antenna simulations can predict the direction-dependent nonlinear response of active arrays~\cite{hausmair_prediction_2017, dhar_reflection-aware_2018,fager_linearity_2019}. In analog and hybrid beamforming transmitters, including multibeam systems, a shared digital signal drives multiple PAs that experience different beam- and frequency-dependent antenna loads. The resulting AM--AM and AM--PM responses can vary across the array. The effectiveness of a common or beam-dependent DPD depends on the consistency of the element responses, the array architecture, and the radiation directions selected for linearization~\cite{capelli-mouvand_5g_2021,holzman_use_2013}.

DPD under antenna VSWR variation has received increasing attention in PA linearization and active-array research. Reflected-wave-dependent memory and nonlinear cross terms have been included in low-complexity DPD models for handset PAs~\cite{wang_digital_2022}. Adaptive approaches can combine a frequency- and bandwidth-dependent nominal model with a compact correction that is updated as the operating band and VSWR change~\cite{li_adaptive_2025}. For beamforming transmitters, multi-target DPD can widen the angular region over which out-of-band emissions are suppressed~\cite{luo_linearization_2021}. A VSWR-resilient PA that maintains similar large-signal responses across the array reduces element-to-element nonlinear variation and, in turn, the adaptation range, model complexity, and recalibration burden of shared or beam-dependent DPD.

\begin{figure}[t]
\centering
\includegraphics[width=0.9\linewidth]{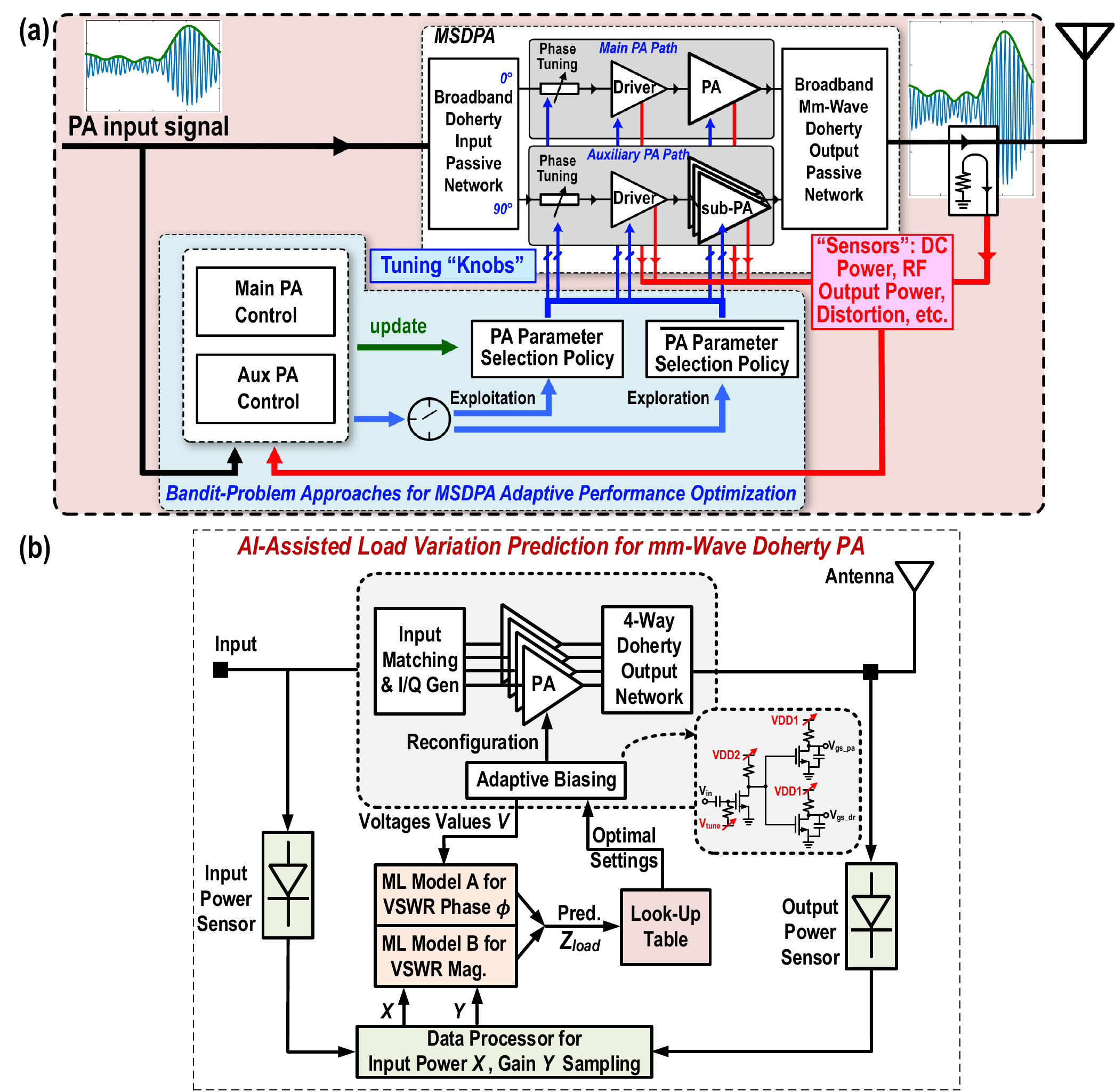}
\caption{AI-assisted adaptation methods for VSWR-resilient PAs. (a) In-field mixed-signal Doherty PA optimization using sensed PA responses and bandit- and reinforcement-learning-based control~\cite{xu_-field_2020, wang_super-resolution_2019, wang_artificial-intelligence_2019}. (b) Complex-load prediction and state selection using input/output power sensing~\cite{svelto_ai-assisted_2025}.}
\label{fig:ai_vswr_pa_flow}
\vspace{-1.5em}
\end{figure}

Future wireless links promise extremely low latency~\cite{wang_millimeter-wave_2021}. In large arrays operating in dynamic and partially unknown environments, short beam-update intervals make rapid PA adaptation increasingly important. The controller must infer a suitable PA state from limited sensor observations without interrupting normal transmitter operation or introducing excessive calibration overhead. Beyond DPD-based linearization, AI/ML-assisted \emph{in-situ} computation can support both online optimization of the PA operating state and complex-load prediction for LUT-based state selection. An example of an AI-assisted mixed-signal Doherty PA is shown in Fig.~\ref{fig:ai_vswr_pa_flow}(a)~\cite{wang_super-resolution_2019, wang_artificial-intelligence_2019,xu_-field_2020}. Sensed PA responses can support bandit- and reinforcement-learning-based updates of the PA drive and bias settings, while prior knowledge of the PA response can reduce the number of states evaluated during adaptation. This reduction in the search space is particularly important when the available update time is limited by beam switching or changing antenna VSWR conditions. The complementary approach in Fig.~\ref{fig:ai_vswr_pa_flow}(b) uses input/output power sensing and the applied PA states to estimate the magnitude and phase of the complex load and select a pre-characterized LUT entry~\cite{svelto_ai-assisted_2025}. These studies demonstrate the potential of AI/ML-assisted adaptation, while the reported VSWR-recovery results remain primarily based on simulation and prior PA characterization.

Beyond post-silicon adaptation, data-driven methods can also support VSWR-resilient PA development before fabrication~\cite{chu_recent_2026,chu_ai-assisted_2025}. Reusable models obtained from electromagnetic (EM) and circuit data can accelerate active--passive co-design and design migration across operating frequencies and technology nodes~\cite{er_deep_2021, chu_transfer_2024, chu_deep_2025,chu_top-metal-only_2026}. AI/ML-assisted design workflows can automate the broad exploration of inherently VSWR-resilient PA cores together with compact, low-loss passive matching networks. The resulting candidates must still undergo circuit simulation based on the process design kit (PDK), full-wave EM analysis, and physical verification before tapeout~\cite{chu_ai-assisted_2026}.

Future array-scalable solutions should combine these techniques across the design, implementation, and control layers. VSWR-resilient PA hardware should first be compact and maintain efficient operation over the expected antenna VSWR region. Low-overhead sensing and reconfiguration can then address residual load- and temperature-dependent variations, while shared or beam-dependent DPD corrects the residual nonlinear distortion under modulated-signal operation. Intelligent control can coordinate load estimation, state selection, and parameter updates. This division of functions can reduce the dimensionality and update rate of the array-control problem and provide a path toward predictable and energy-efficient operation in future communication and sensing arrays.

\section{Conclusion}

The rapid development of large-scale mm-Wave and cm-Wave phased arrays has introduced new challenges in integrated PA design, implementation, operation, and calibration. The PAs and antenna elements form a coupled interface in which the PA output amplitude and phase determine the radiated array response, while the frequency-, scan-angle-, and element-position-dependent antenna active impedance perturbs the operating condition of each PA. This review discusses the origins and system-level impact of antenna VSWR, establishes how the interaction between $\Gamma_{\mathrm{ant}}$ and $S_{22}$ affects delivered power, transmitted phase, and RIMD, and reviews recent integrated silicon PA techniques and circuit demonstrations. The analysis motivates PA designs that achieve simultaneous output matching and loadline matching (SOLM). Future research should target ultra-compact implementations, VSWR resilience with PBO efficiency enhancement, higher power density, and \emph{in-situ} adaptation for array-scalable integrated PAs. Continued progress in these directions will further advance mm-Wave and cm-Wave PAs for future communication and sensing systems.

\bibliographystyle{IEEEtran}
\bibliography{references}

\end{document}